\documentclass[12pt,preprint]{aastex}

\begin{document}

\title{Searching for a Stochastic Background of Gravitational Waves with LIGO}


\author{B.~Abbott\altaffilmark{12},
R.~Abbott\altaffilmark{12},
R.~Adhikari\altaffilmark{12},
J.~Agresti\altaffilmark{12},
P.~Ajith\altaffilmark{2},
B.~Allen\altaffilmark{42},
R.~Amin\altaffilmark{16},
S.~B.~Anderson\altaffilmark{12},
W.~G.~Anderson\altaffilmark{42},
M.~Araya\altaffilmark{12}
% ...
}

\affil{The LIGO Scientific Collaboration, http://www.ligo.org}


% ...
\altaffiltext {2}{Albert-Einstein-Institut, Max-Planck-Institut f\"ur Gravitationsphysik, D-30167 Hannover, Germany}
% ...
\altaffiltext {12}{LIGO - California Institute of Technology, Pasadena, CA  91125, USA}
% ...
\altaffiltext {16}{Louisiana State University, Baton Rouge, LA  70803, USA}
% ...
\altaffiltext {42}{University of Wisconsin-Milwaukee, Milwaukee, WI  53201, USA}
% ...

\begin{abstract}
The Laser Interferometer Gravitational-wave Observatory (LIGO) has performed
the fourth science run, S4, with significantly improved interferometer
sensitivities with respect to previous runs. Using data acquired during this 
science run, we place a limit on the amplitude of a stochastic
background of gravitational waves. For a frequency independent spectrum, 
the new limit is $\Omega_{\rm GW} < 6.5 \times 10^{-5}$. This is currently the
most sensitive result in the frequency range 51-150 Hz, with a factor of 13
improvement over the previous LIGO result. We discuss complementarity of
the new result with other constraints on a stochastic background
of gravitational waves, and we investigate implications of
the new result for different models of this background.
\end{abstract}

\thispagestyle{myheadings}
\markboth{}{LIGO-P060012-05-D}


\section{Introduction}

A stochastic background of gravitational waves (GWs) is expected to arise as 
a superposition of a large number of unresolved sources, from different
directions in the sky, and with different polarizations. It 
is usually described in terms of the GW spectrum:
\begin{equation}
\Omega_{\rm GW}(f) = \frac{f}{\rho_c} \; \frac{d \rho_{\rm GW}}{d f}\,,
\end{equation}
where $d\rho_{\rm GW}$ is the energy density of gravitational radiation
contained in the frequency range $f$ to $f+df$ \citep{allenromano},
$\rho_c$ is the critical energy density of the Universe, and $f$ is
frequency (for an alternative and equivalent definition of $\Omega_{\rm GW}(f)$
see, for example, \citep{baskaran}). 
In this paper, we will focus on power-law GW spectra:
\begin{equation}
\Omega_{\rm GW}(f) = \Omega_{\alpha} \Bigg( \frac{f}{100 {\rm \; Hz}} 
\Bigg)^{\alpha}.
\label{template}
\end{equation}
Here, $\Omega_{\alpha}$ is the amplitude corresponding to the spectral
index $\alpha$. In particular, $\Omega_0$ denotes the amplitude of the
frequency-independent GW spectrum.

Many possible sources of stochastic GW 
background have been proposed and several experiments have 
searched for it (see \citep{maggiore,allen} for reviews). Some of 
the proposed theoretical models are cosmological in nature, such as 
the amplification of quantum vacuum 
fluctuations during inflation \citep{PA1}, \citep{PA2}, \citep{star}, 
pre-big-bang models \citep{pbb}, \citep{pbbrep}, \citep{BMU}, 
phase transitions \citep{PT1}, \citep{PT2}, and cosmic strings \citep{CS1}, 
\citep{CS2}, \citep{CS3}. Others are astrophysical in nature, such
as rotating neutron stars \citep{RFP}, 
supernovae \citep{coward} or low-mass X-ray binaries \citep{cooray}.

A number of experiments have been used to constrain the spectrum
of GW background at different frequencies. Currently, the most stringent
constraints arise from large-angle correlations in the cosmic
microwave background (CMB) \citep{cobe1,cobe2}, from the 
arrival times of millisecond pulsar signals \citep{pulsar},
from Doppler tracking of the Cassini spacecraft \citep{doppler}, and
from resonant bar GW detectors, such as Explorer and Nautilus \citep{bars}.
An indirect bound can be placed on the total energy carried by gravitational 
waves at the time of the Big-Bang Nucleosynthesis (BBN) using the BBN 
model and observations 
\citep{BBN,maggiore,allen}. Similarly, \citep{smith} used 
the CMB and matter spectra to constrain the total energy density of
gravitational waves at the time of photon decoupling.

Ground-based interferometer networks can directly measure the GW strain
spectrum in the frequency band 10 Hz - few kHz, by searching for correlated
signal beneath uncorrelated detector noise. LIGO has built three 
power-recycled Michelson interferometers, with a 
Fabry-Perot cavity in each orthogonal arm. They are located at two sites,
Hanford, WA, and Livingston Parish, LA. There are two collocated 
interferometers at the WA site: H1, with 4km long arms, and H2, with 2km 
arms. The LA site contains L1, a 4km interferometer, similar in design to H1.
The detector configuration and performance 
during LIGO's first science run (S1) was described in \citep{S1}. The data
acquired during that run was used to place an upper limit of 
$\Omega_0 < 44.4$ on the amplitude of a frequency independent
GW spectrum, in the frequency band 40-314 Hz \cite{S1stoch}.
This limit, as well as the rest of this paper, assumes the present value of 
the Hubble parameter $H_0 = 72$ km/s/Mpc or, equivalently, 
$h_{100} = H_0 / (100 {\rm \; km/s/Mpc}) = 0.72$ \citep{hubble}.
The most recent bound on the amplitude of the frequency independent 
GW spectrum from LIGO is based on the
science run S3: $\Omega_0 < 8.4 \times 10^{-4}$ 
for a frequency-independent spectrum in the 69-156 Hz band~\citep{S3paper}. 
\begin{figure}
\epsscale{.80}
\plotone{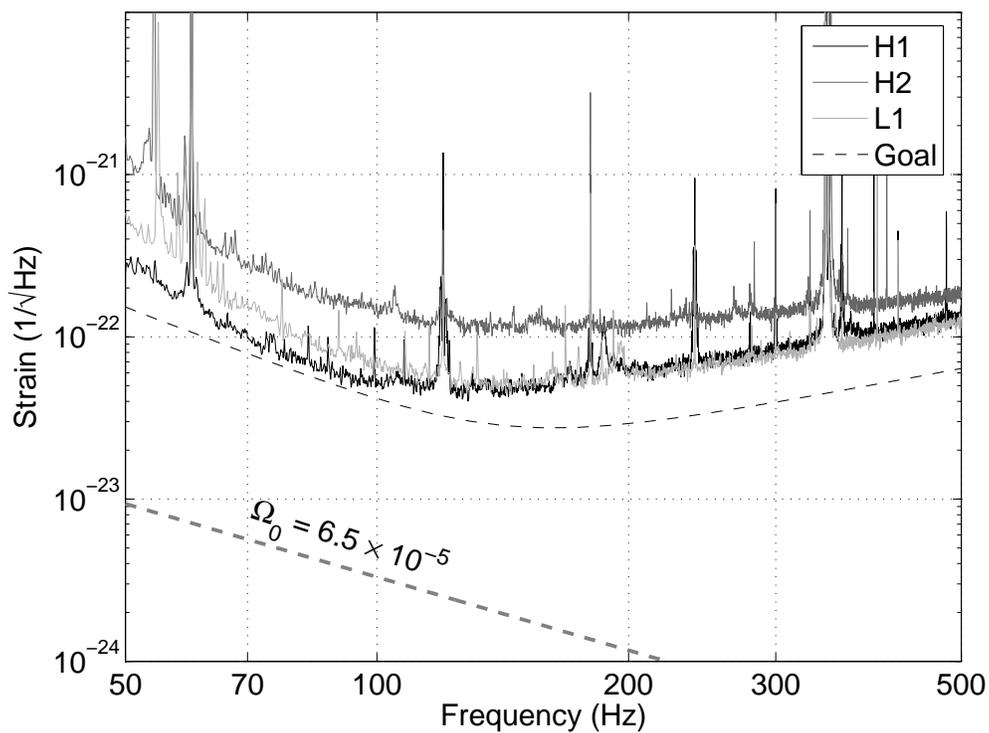}
\caption{Typical strain amplitude spectra of 
LIGO interferometers during the science run S4 (solid curves top-to-bottom at
70 Hz: H2, L1, H1). The black dashed curve is the LIGO sensitivity goal.
The gray dashed curve is the strain amplitude spectrum corresponding to the
limit presented in this paper for the frequency-independent GW spectrum
$\Omega_0 < 6.5 \times 10^{-5}$.}
\label{strain}
\end{figure}

In this paper, we report much improved 
limits on the stochastic GW background around 100 Hz, using
the data acquired during the LIGO science run S4, which took place between
February 22, 2005 and March 23, 2005. The sensitivity of the interferometers
during S4, shown in Figure \ref{strain}, was significantly better as compared 
to S3 (by a factor $10\times$ at certain frequencies), 
which leads to an order of magnitude improvement 
in the upper limit on the amplitude of the stochastic GW background:
$\Omega_0 < 6.5 \times 10^{-5}$ for a frequency-independent spectrum 
over the 51-150 Hz band. 

This limit is beginning to probe some models of the stochastic
GW background. As examples, we investigate the implications of this
limit for cosmic strings models and for pre-big-bang models of
the stochastic gravitational radiation. In both cases, the new LIGO result
excludes parts of the parameter space of these models. 

The organization of this paper is as follows. 
In Section 2 we review the analysis procedure and present the results 
in Section 3. In Section 4, we discuss some of the implications
of our results for models of a stochastic GW background, as well
as the complementarity between LIGO and other experimental constraints on
a stochastic GW background. We conclude with future prospects in Section 5. 

\section{Analysis}

\subsection{Cross-Correlation Method}

\begin{figure}
\epsscale{.80}
\plotone{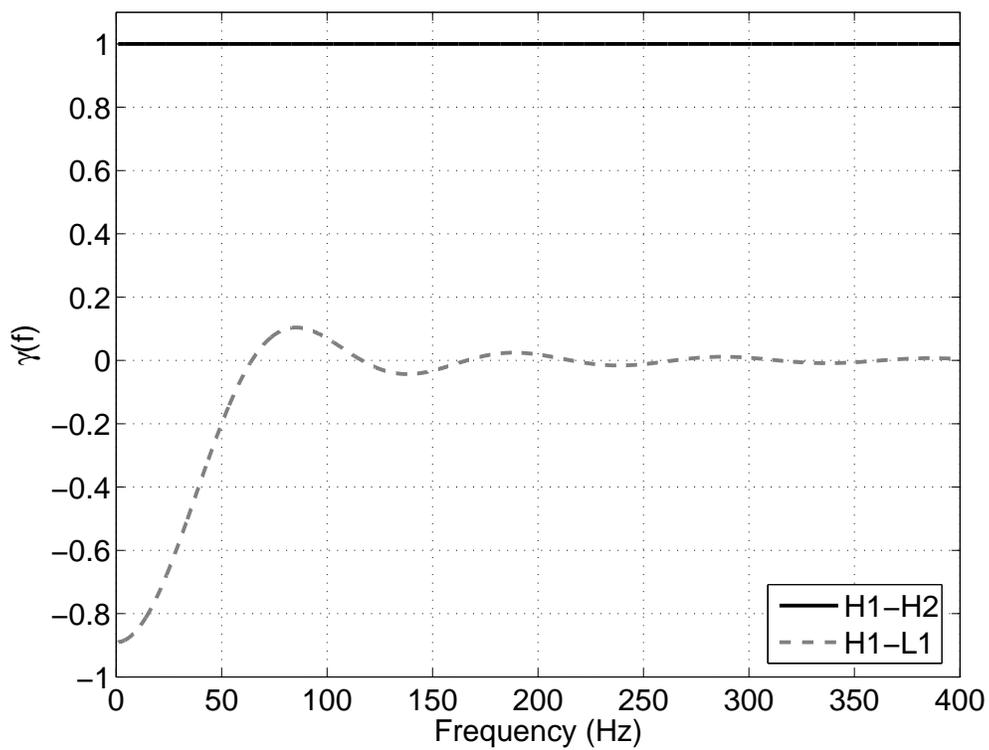}
\caption{Overlap reduction function for the Hanford-Hanford pair
(black solid) and for the Hanford-Livingston pair (gray dashed).}
\label{overlap}
\end{figure}

The cross-correlation method for searching for a stochastic GW background with 
pairs of ground-based interferometers is described in \citep{allenromano}. We 
define the following cross-correlation estimator:
\begin{eqnarray}
Y & = & \int_{0}^{+\infty } df \; Y(f) \nonumber \\
& = & \int_{-\infty }^{+\infty } df \int_{-\infty }^{+\infty } df' \;
\delta_T (f-f') 
\; \tilde{s}_1(f)^{*} \; \tilde{s}_2(f') \; \tilde{Q}(f')\,,
\label{ptest}
\end{eqnarray}
where $\delta_T$ is a finite-time approximation to the Dirac delta function,
$\tilde{s}_1$ and $\tilde{s}_2$ are the Fourier transforms of the
strain time-series of two interferometers, and $\tilde{Q}$ is
a filter function. Assuming that the detector noise
is Gaussian, stationary, uncorrelated between the two interferometers,
and much larger than the GW signal, the
variance of the estimator $Y$ is given by:
\begin{eqnarray}
\sigma_Y^2 & = & \int_0^{+\infty} df \; \sigma_Y^2(f) \nonumber \\
& \approx & \frac{T}{2} \int_0^{+\infty} df P_1(f) P_2(f)
\mid \tilde{Q}(f) \mid^2\,,
\label{sigma}
\end{eqnarray}
where $P_i(f)$ are the one-sided power spectral densities (PSDs) 
of the two interferometers
and $T$ is the measurement time. Optimization of the signal-to-noise
ratio leads to the following form of the 
optimal filter \citep{allenromano}:
\begin{equation}
\label{optfilt1}
\tilde{Q}(f) = \mathcal{N} \; \frac{\gamma(f) 
S_{\rm GW}(f)}{P_1(f) P_2(f)} \; ,
\end{equation}
where
\begin{equation}
S_{\rm GW}(f) = \frac{3 H_0^2}{10 \pi^2} \; 
\frac{\Omega_{\rm GW}(f)}{f^3} \; ,
\label{optfilt2}
\end{equation}
and $\gamma(f)$ is the overlap reduction function, arising from the
different locations and orientations of the two interferometers. As shown
in Figure \ref{overlap}, the identical antenna patterns of the 
collocated Hanford interferometers imply $\gamma(f) = 1$. 
For the Hanford-Livingston pair the overlap reduction is significant 
above 50 Hz. 
In Equations \ref{optfilt1} and \ref{optfilt2}, $S_{\rm GW}(f)$ is 
the strain power spectrum of the stochastic GW background to be searched. 
Assuming a power-law template GW spectrum with index $\alpha$
(see Equation \ref{template}),
the normalization constant $\mathcal{N}$ in Equation \ref{optfilt1} 
is chosen such that $<Y> = \Omega_{\alpha} T$.

In order to deal with data non-stationarity, and for purposes of computational
feasibility, the data for an interferometer pair are divided into 
many intervals of equal
duration, and $Y_I$ and $\sigma_{Y_I}$ are calculated for each interval $I$. 
The data in each interval are decimated from 16384 Hz to 1024 Hz and 
high-pass filtered with a 40 Hz cut-off. They are also Hann-windowed to
avoid spectral leakage from strong lines present in the data. Since 
Hann-windowing effectively reduces the interval length by 50\%, the
data intervals are overlapped by 50\% to recover the original signal-to-noise
ratio. The effects of windowing are taken into account as discussed in 
\cite{S1stoch}.

The PSDs for each interval (needed for the calculation of $Q_I(f)$ and of 
$\sigma_{Y_I}$) are calculated using the two neighboring intervals.
This approach avoids a bias that would otherwise exist due to a non-zero
covariance between the cross-power $Y(f)$ and the
power spectra estimated from the same data \citep{bendat}. 
It also allows for a stationarity cut, which we will describe
in more detail below. 

We consider two interval durations and frequency resolutions:
\begin{itemize}
\item 60-sec duration with $1/4$ Hz resolution: the PSDs are calculated
by averaging 58 50\% overlapping periodograms (based on the two neighboring 
60-sec intervals) in Welch's modified periodogram method.
\item 192-sec duration with $1/32$ Hz resolution: the PSDs are calculated
by averaging 22 50\% overlapping periodograms (based on the two neighboring 
192-sec intervals) in Welch's modified periodogram method.
\end{itemize}

As we will discuss below, the 60-sec intervals allow better sensitivity to
noise transients and are better suited for data-stationarity cuts, while 
the 192-sec intervals allow higher frequency resolution of the power and 
cross-power spectra and are better suited for removing sharp lines
from the analysis.

The data for a given interval $I$ are Fourier transformed and rebinned
to the frequency resolution of the optimal filter 
to complete the calculation of $Y_I$ (Eq. \ref{ptest}). 
Both the PSDs and the Fourier transforms of the data are calibrated using
interferometer response functions, determined for every minute of data using a 
measurement of the interferometer response to a sinusoidal calibration force. 
To maximize the signal-to-noise ratio, the intervals
are combined by performing a weighted average 
(with weights $1/\sigma^2_{Y_I}$),
properly accounting for the 50\% overlapping as discussed in \citep{lazzrom}.

\subsection{Identification of Correlated Instrumental Lines}

\begin{figure}
\epsscale{.60}
\plotone{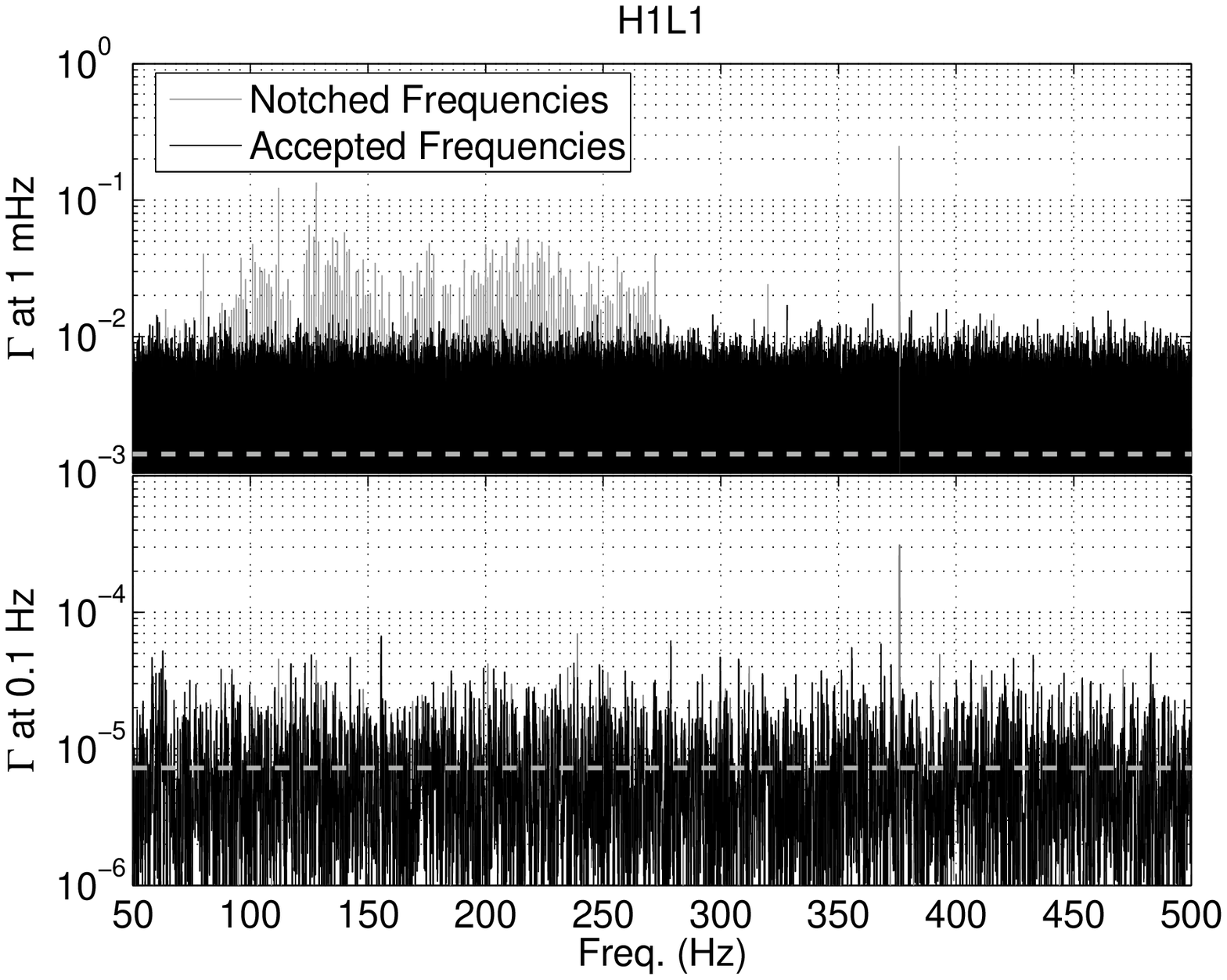}
\plotone{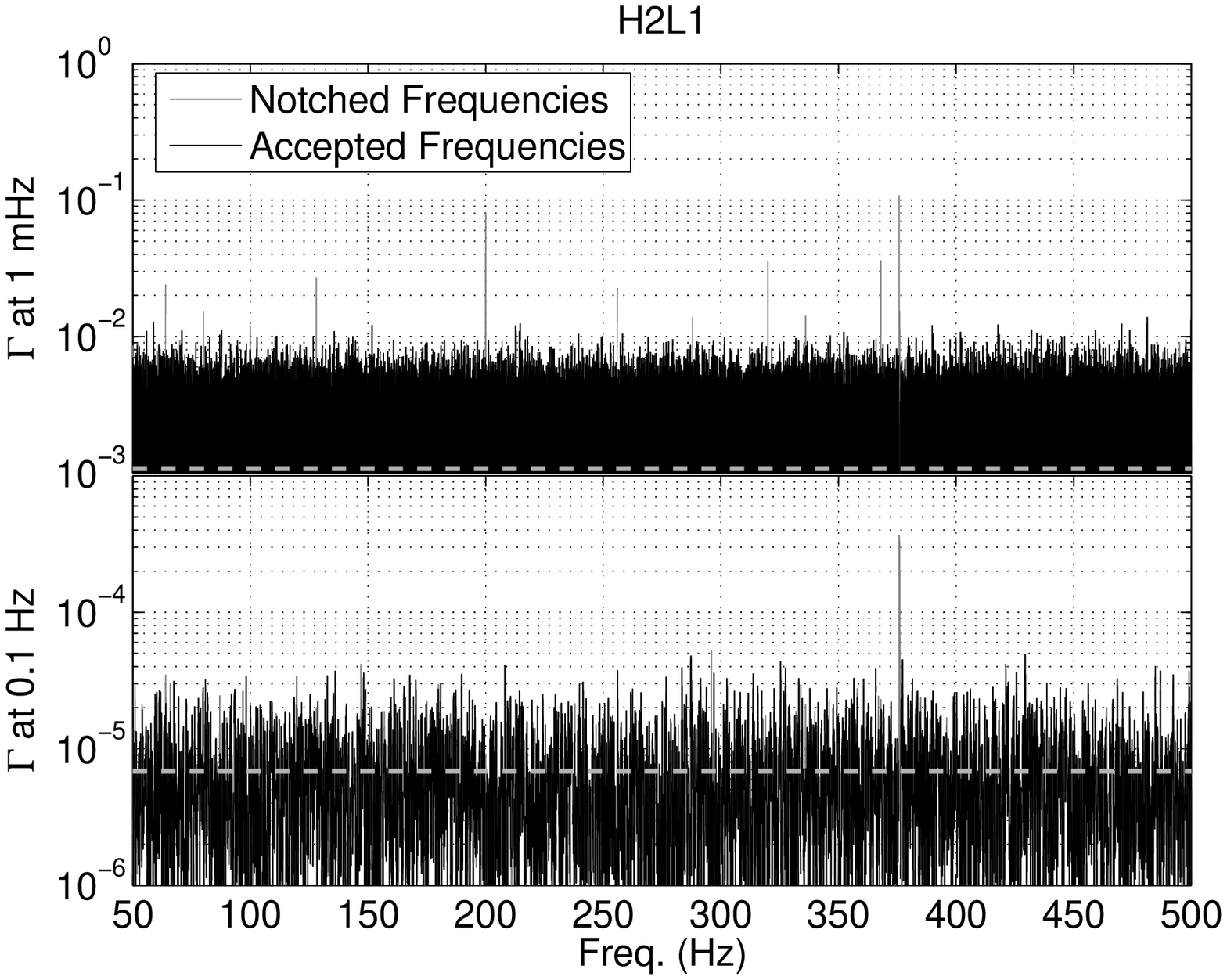}
\caption{Coherence calculated for the H1L1 pair (top) and for the H2L1
pair (bottom) over all of S4 data for 1 mHz resolution and 100 mHz resolution.
The horizontal dashed lines indicate $1/N_{avg}$ - the expected level of 
coherence after averaging over $N_{avg}$ time-periods with uncorrelated 
spectra. The line at 376 Hz is one of the simulated pulsar lines.}
\label{coherence}
\end{figure}

The results of this paper are based on the Hanford-Livingston interferometer
pairs, for which the broadband instrumental correlations are
minimized. Nevertheless, it is still necessary to investigate if there
are any remaining periodic instrumental correlations. We do this by 
calculating the coherence over the whole S4 run. The coherence is defined as:
\begin{equation}
\Gamma(f) = \frac{|s_1^*(f) s_2(f)|^2}{P_1(f) P_2(f)}.
\end{equation}
The numerator is the square of the cross-spectral density (CSD) between the two
interferometers, and the denominator contains the two power spectral densities
(PSDs). We average the CSD and the PSDs over the whole run
at two different resolutions: 1 mHz and 100 mHz. Figure \ref{coherence} shows
the results of this calculation for the H1L1 pair and for the H2L1 pair.

\begin{figure}
\epsscale{.60}
\plotone{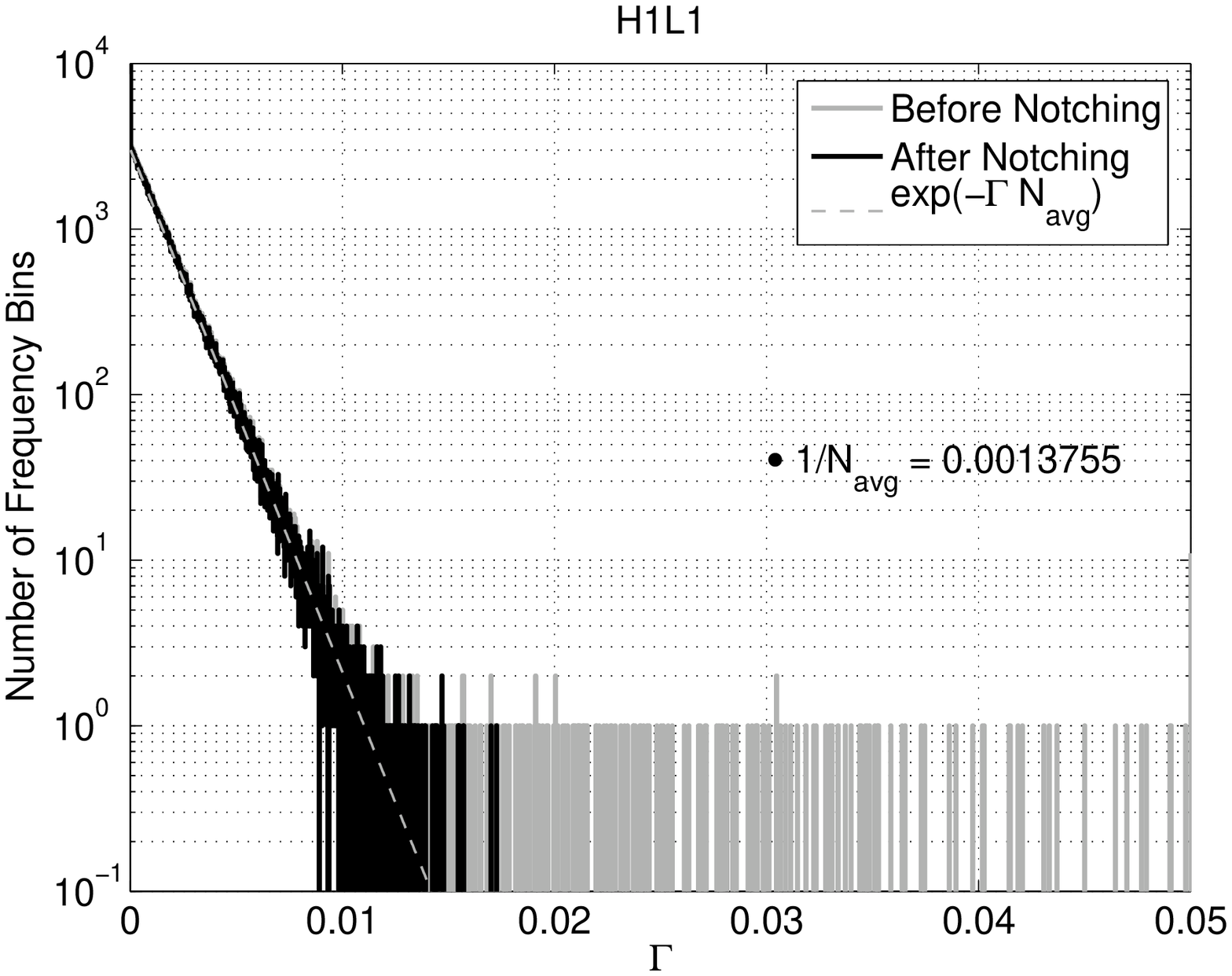}
\plotone{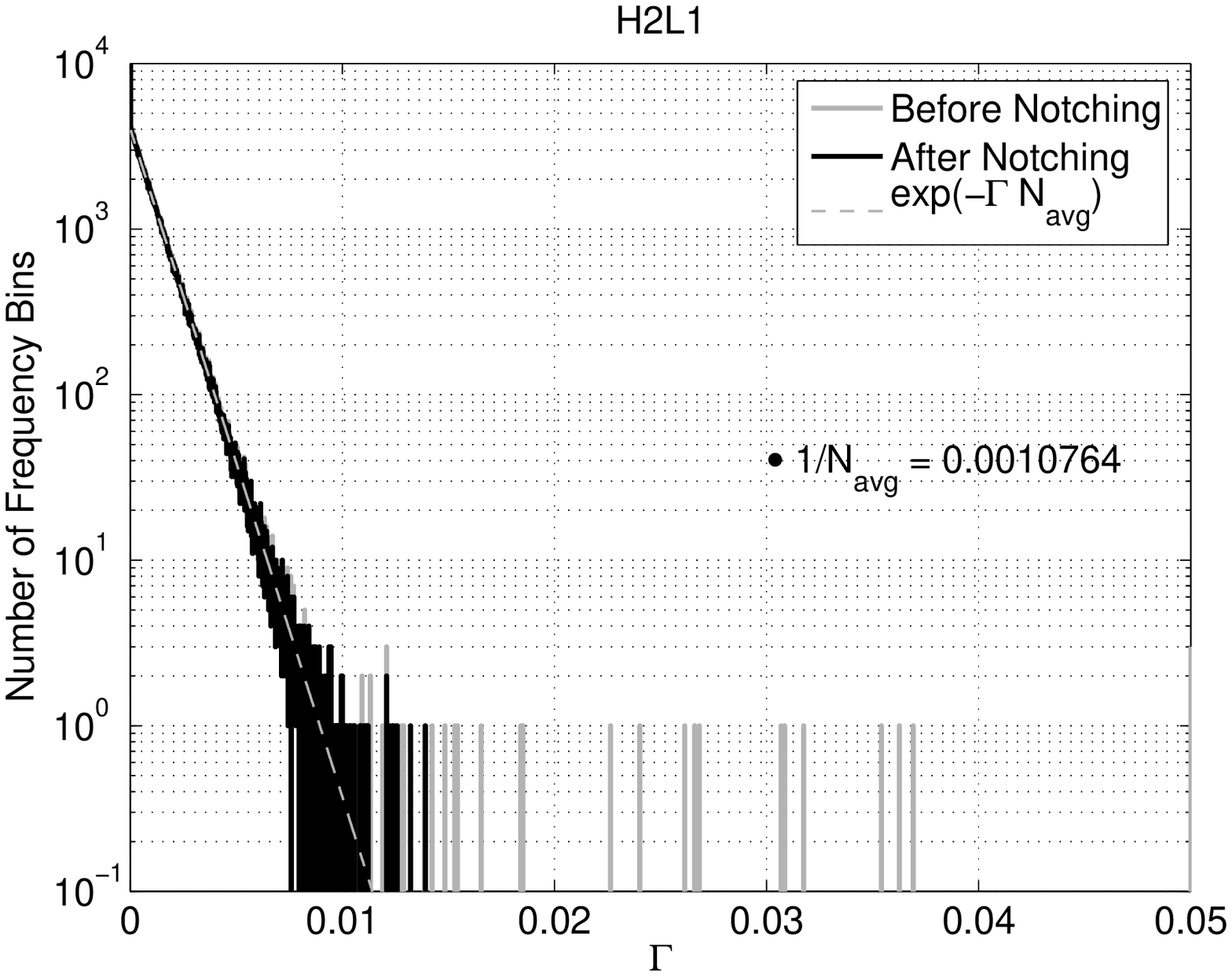}
\caption{Histogram of the coherence for H1L1 (top) and H2L1 (bottom) at 1 mHz
resolution follows the expected exponential distribution, with exponent
coefficient $N_{avg}$ (the number of time-periods
over which the average is made).}
\label{cohfit}
\end{figure}

At 1 mHz resolution, a forest of sharp 1 Hz harmonic lines can 
be observed. These lines were likely caused by
the sharp ramp of a one-pulse-per-second signal, injected into the 
data acquisition system to synchronize it with the Global Positioning System 
(GPS) time reference. After the S4 run ended, the sharp ramp signal was
replaced by smooth sinusoidal signals, with the goal of significantly
reducing the 1 Hz harmonic lines in future LIGO data runs. 
In addition to the 1 Hz lines, the 1 mHz coherence plots in Figure 
\ref{coherence} also include some of the simulated pulsar lines, 
which were injected into the differential-arm servo of the interferometers
by physically moving the mirrors.
Both the 1 Hz harmonics and the simulated pulsar lines can be removed 
in the final analysis, and we will discuss this
further in Section 3. Figure \ref{cohfit} shows that the histogram
of the coherence at 1 mHz resolution follows the expected exponential
distribution, if one ignores the 1 Hz harmonics and the simulated pulsar lines.

\subsection{Data Quality Cuts}

In our analysis, we include time periods during which both interferometers 
are in low-noise, science mode. We exclude:
\begin{itemize}
\item Time-periods when digitizer signals saturate.
\item 30-sec intervals prior to each lock loss. These intervals 
are known to be particularly noisy.
\end{itemize}

\begin{figure}
\epsscale{.80}
\plotone{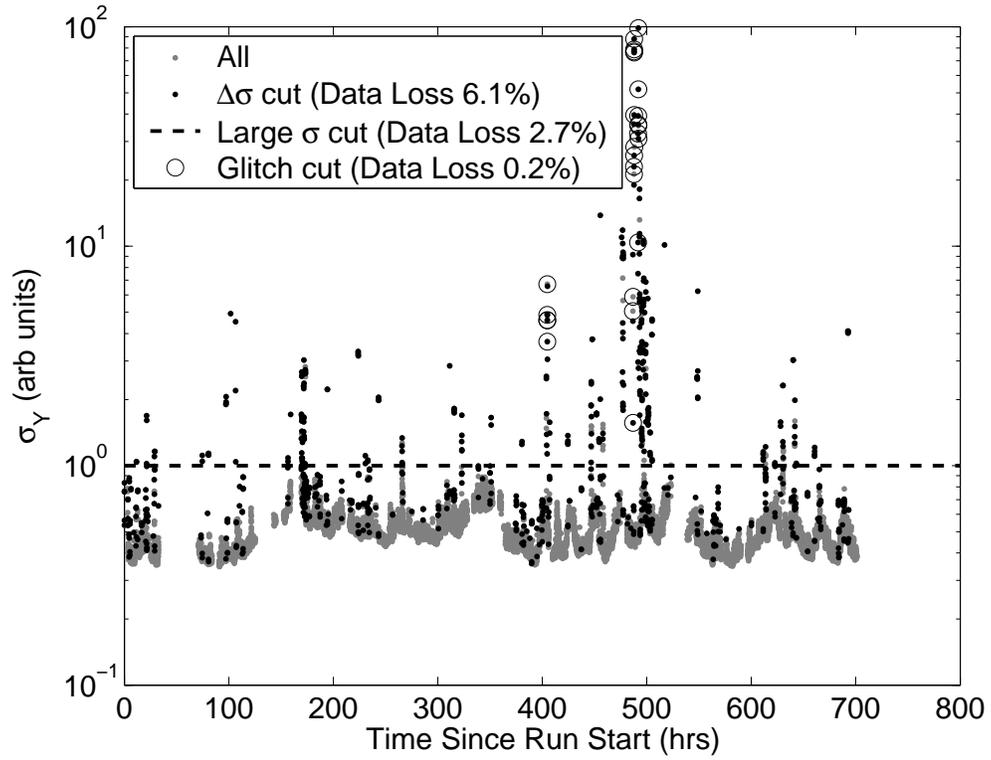}
\caption{Trend of $\sigma_{Y_I}$ over the whole S4 run for the 192-sec
intervals of H1L1 pair. The dashed line denotes the large $\sigma$ cut: 
segments lying above this line are removed from analysis. 
Note the daily variation in the sensitivity of this pair.}
\label{badGPSTimes}
\end{figure}

We then proceed to calculate $Y_I$ and $\sigma_{Y_I}$ for each interval $I$,
and define three data-quality cuts. First,
we reject intervals known to contain large glitches in one 
interferometer. These intervals were identified by
searching for discontinuities in the PSD trends over the whole S4 run.
Second, we reject intervals for which $\sigma_{Y_I}$ is 
anomalously large. In particular, for the 192-sec analysis, we require
$\sigma_{Y_I} < 1$ sec for the H1L1 pair, and $\sigma_{Y_I} < 2$ sec
for the H2L1 pair (recall that $Y$ is normalized such that 
$<Y> = \Omega_{\alpha} T$, with $T=192$ sec in this case). 
The glitch cut and the large-sigma cut largely 
overlap, and are designed to
remove particularly noisy time-periods from the analysis. Note, also, that
due to the weighting with $1/\sigma^2_{Y_I}$, the contribution of these
intervals to the final result would be suppressed, but we reject them
from the analysis nevertheless. Third, we reject
the intervals for which 
$\Delta\sigma=|\sigma_{Y_I} - \sigma_{Y_I}'|/\sigma_{Y_I} > \zeta$. Here,
$\sigma_{Y_I}$ is calculated using the two intervals neighboring 
interval $I$, and $\sigma_{Y_I}'$ is calculated using the interval $I$ 
itself. The optimization of threshold $\zeta$ is
discussed below. The goal of this cut is to capture noise-transients
in the data, and reject them from the analysis.
Figure \ref{badGPSTimes} shows the impact of these
cuts for the H1L1 pair, analyzed with 192-sec segments, 1/32 Hz resolution,
and with $\zeta = 0.3$. This Figure also shows daily variation in the
sensitivity to stochastic GW background, arising from the variation
in the strain sensitivity of the interferometers, which is typically
worse during the week-days than during the weekends or nights. 

\begin{figure}
\epsscale{.80}
\plotone{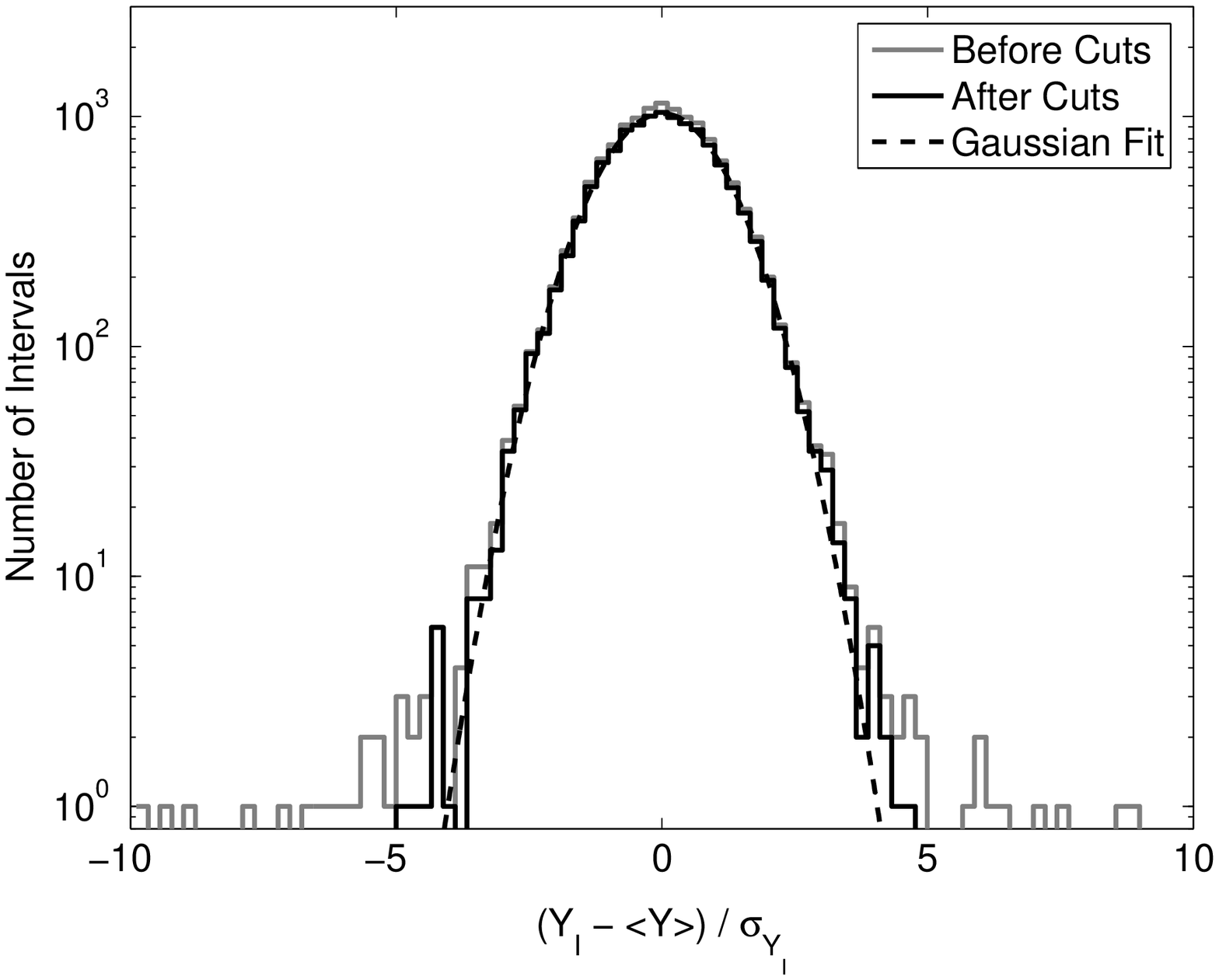}
\caption{Distribution of residuals for the H1L1 pair with 192-sec segments: 
all data are shown in gray,
data that passes data quality cuts are shown in black, and the Gaussian
fit to the black histogram is shown as a dashed curve.}
\label{outlierfit}
\end{figure}

Figure \ref{outlierfit} shows the distribution of the residuals for the same
analysis. For a given
interval $I$, the residual is defined as 
\begin{equation}
\frac{Y_I - <Y>}{\sigma_{Y_I}}.
\end{equation}
Note that the data quality cuts remove outliers from the residual
distribution, hence making the data more stationary. After the cuts,
the Kolmogorov-Smirnov test indicates that the residual distribution is
consistent with a Gaussian, for both 
H1L1 and H2L1 analyses with 192-sec intervals, 
1/32 Hz resolution, and $\zeta=0.3$.

\section{Results}

\subsection{New Upper Limit}

We performed a ``blind'' analysis for the H1L1 and the H2L1 pairs
with 60-sec intervals, $1/4$ Hz resolution, and $\zeta=0.2$. 
To avoid biasing the results, all data-quality cuts were 
defined based on studies done with a 0.1 sec time-shift
between the two interferometers in a pair. Such a time-shift
removes any GW correlations, without significantly affecting the 
instrumental noise
performance. After the data quality cuts were finalized,
we made one last pass through the data, with zero time-shift, and
obtained the final results of the blind analysis. 

The results from the blind analysis for the frequency-independent 
template spectrum
($\alpha=0$) are listed in the first row of Table \ref{table1} for H1L1
and in the first row of Table \ref{table2} for H2L1. These results
show no evidence of a stochastic GW background. After completing the 
blind analysis, we discovered that the instrumental 1 Hz harmonic
lines, discussed in Section 2.2, are correlated between the two sites. 
We felt compelled on scientific grounds not to
ignore these correlations, even though they had been discovered after our
initial, blind, analysis was complete.

\begin{figure}
\epsscale{.80}
\plotone{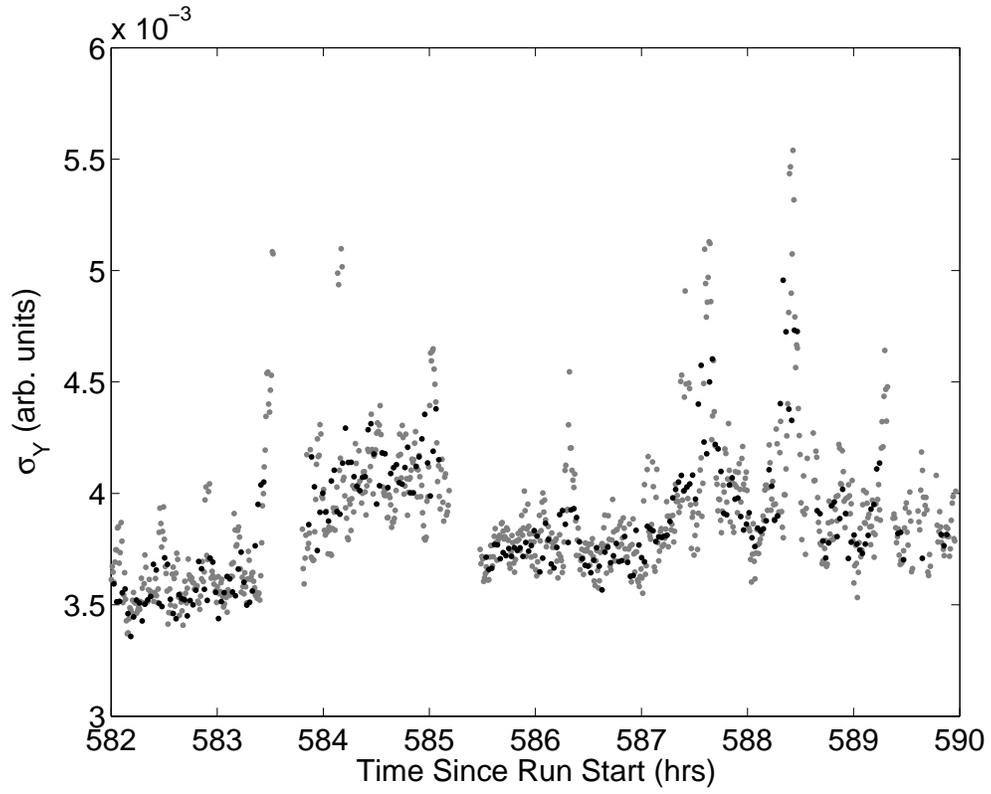}
\caption{Trend of $\sigma_Y$ for the 60-sec analysis (gray) and 192-sec
analysis (black) over a short period of time. The two bands were scaled 
to overlap. Note that the gray band is wider, indicating that the 60-sec
analysis is more sensitive to noise variations.}
\label{60vs192}
\end{figure}

In order to remove from our results any possible influence of the correlated 
lines, we repeated our analysis with refined frequency resolution 
of $1/32$ Hz. We increased the interval length from 60-sec to 192-sec, which
implies that the PSDs are estimated by averaging 22 50\% overlapping 
periodograms. 
These changes allowed us to exclude the 1 Hz harmonics from the analysis while 
losing only $\sim 3$\% of the bandwidth. The drawback, however, was
that the 192-sec analysis was less able to identify and exclude 
the noise-transients
than the 60-sec analysis, as shown in Figure \ref{60vs192}. As a result,
the $\Delta\sigma$ cut was retuned for the 192-sec analysis after the 60-sec
analysis with zero time-shift was completed.

We examined several methods of defining the data quality cuts for 
the 192-sec analysis. First, 
we calculated the $\Delta\sigma$ cut in the 60-sec analysis with $\zeta=0.2$
(along with large sigma and glitch cuts), and then declared 192-sec intervals
``bad'' if they overlapped with the ``bad'' 60-sec intervals. This approach
has the advantage that the 60-sec analysis is more sensitive to
transients, and the disadvantage that a significant fraction of the data
is rejected. The results of this approach are summarized in the second row
of Tables \ref{table1} and \ref{table2}. 
Second, we defined all cuts directly on 192-sec intervals. We
varied the value of $\zeta$ (0.2, 0.3, and 0.4) and selected $\zeta=0.3$ as 
the optimal parameter choice. This choice represents the best compromise
between data loss and data quality: it rejects the least amount of data, while 
still excluding all residual outliers and preserving the Gaussianity of
the data.
The last three rows of Tables \ref{table1} and \ref{table2} summarize 
the results for the three values of $\zeta$. 
\begin{table*}[!th]
\centering
\begin{tabular}{|r|r|r|r|r|r|}
\hline
Analysis & $\zeta$ & Fraction of Data & $\Omega_0 (\times 10^{-5})$ & $\Omega_0 (\times 10^{-5})$ & 90\% UL \\
 & & Excluded & before notching & after notching & $(\times 10^{-5})$ \\
 & & & 50-500 Hz & 51-150 Hz & \\ 
\hline
60-sec (blind) & 0.2 & 7.1\% & $6.4 \pm 4.3$ & - & 13.0 \\
192-sec & conversion & 21.0\% & $3.4 \pm 4.9$ & $-2.4 \pm 5.0$ & 7.0 \\
192-sec & 0.2 & 10.9\% & $6.1 \pm 4.7$ & $0.1 \pm 4.8$ & 8.0 \\
192-sec & 0.3 & 6.5\% & $4.8 \pm 4.6$ & $-0.7 \pm 4.7$ & 7.3 \\
192-sec & 0.4 & 5.0\% & $4.8 \pm 4.6$ & $-0.7 \pm 4.7$ & 7.3 \\
\hline
\end{tabular}
\caption{H1L1 results, for $h_{100}=0.72$ and for a frequency-independent
template spectrum ($\alpha = 0$). 
The first two columns define the analysis (interval duration and 
$\Delta\sigma$ cut). For the 192-sec analysis with $\zeta$ denoted as 
``conversion'' (second row), the $\Delta\sigma$ cut was defined using 
the 60-sec analysis listed in the first row: the 192-sec segments are
declared ``bad'' if they overlap with a ``bad'' 60-sec segment.
The third column shows the
fraction of the data lost to data-quality cuts. The fourth column lists
the estimates of $\Omega_0$ for the 50-500 Hz range before notching the 
1 Hz harmonics and the simulated pulsar lines. The fifth column lists the 
estimates of $\Omega_0$ for the $51-150$ Hz range after notching the 1 Hz 
harmonics and the simulated pulsar lines - this frequency
range was determined to be optimal for the combined H1L1+H2L1 analysis 
(see text). The sixth column
gives the 90\% upper limit based on the result in the fifth column. For all
analyses presented here, the distribution of the residual outliers was 
consistent with a Gaussian distribution (c.f. Figure \ref{outlierfit};
the Kolmogorov-Smirnov test statistic
for comparing the two distributions was greater than 85\% in all cases).}
\label{table1}
\end{table*}

\begin{table*}[!th]
\centering
\begin{tabular}{|r|r|r|r|r|r|}
\hline
Analysis & $\zeta$ & Fraction of Data & $\Omega_0 (\times 10^{-5})$ & $\Omega_0 (\times 10^{-5})$ & 90\% UL \\
 & & Excluded & before notching & after notching & $(\times 10^{-5})$ \\
 & & & 50-500 Hz & 51-150 Hz & \\ 
\hline
60-sec (blind) & 0.2 & 4.1\% & $-8.2 \pm 10.6$ & - & 13.0 \\
192-sec & conversion & 10.7\% & $-3.9 \pm 11.9$ & $-7.2 \pm 12.1$ & 16.1 \\
192-sec & 0.2 & 6.1\% & $-1.7 \pm 11.6$ & $-4.2 \pm 11.8$ & 17.1 \\
192-sec & 0.3 & 4.6\% & $-0.6 \pm 11.5$ & $-3.3 \pm 11.7$ & 17.5 \\
192-sec & 0.4 & 3.8\% & $-3.1 \pm 11.5$ & $-6.1 \pm 11.7$ & 15.9 \\
\hline
\end{tabular}
\caption{H2L1 results, for $h_{100}=0.72$ and for a frequency-independent
template spectrum ($\alpha=0$). 
The columns are as in Table \ref{table1}. For all
analyses presented here, the distribution of the residual outliers was 
consistent with a Gaussian distribution (the Kolmogorov-Smirnov test statistic
for comparing the two distributions was greater than 52\% in all cases).}
\label{table2}
\end{table*}

The 192-sec analysis was not performed blindly. However, it agrees with our
blind 60-sec analysis very well, because the contribution of the
correlated 1 Hz harmonic lines is only about 1.5$\sigma$. It is also more 
conservative than the blind analysis 
since the value of the theoretical error is larger due to a smaller
amount of data available in the form of acceptable 
192-sec intervals (as compared to the 60-sec intervals).
It also properly
handles the known instrumental correlations at 1 Hz harmonics. Hence, in
the remaining part of the paper we will focus on the 192-sec analysis
with $\zeta=0.3$. 

\begin{figure}
\epsscale{.80}
\plotone{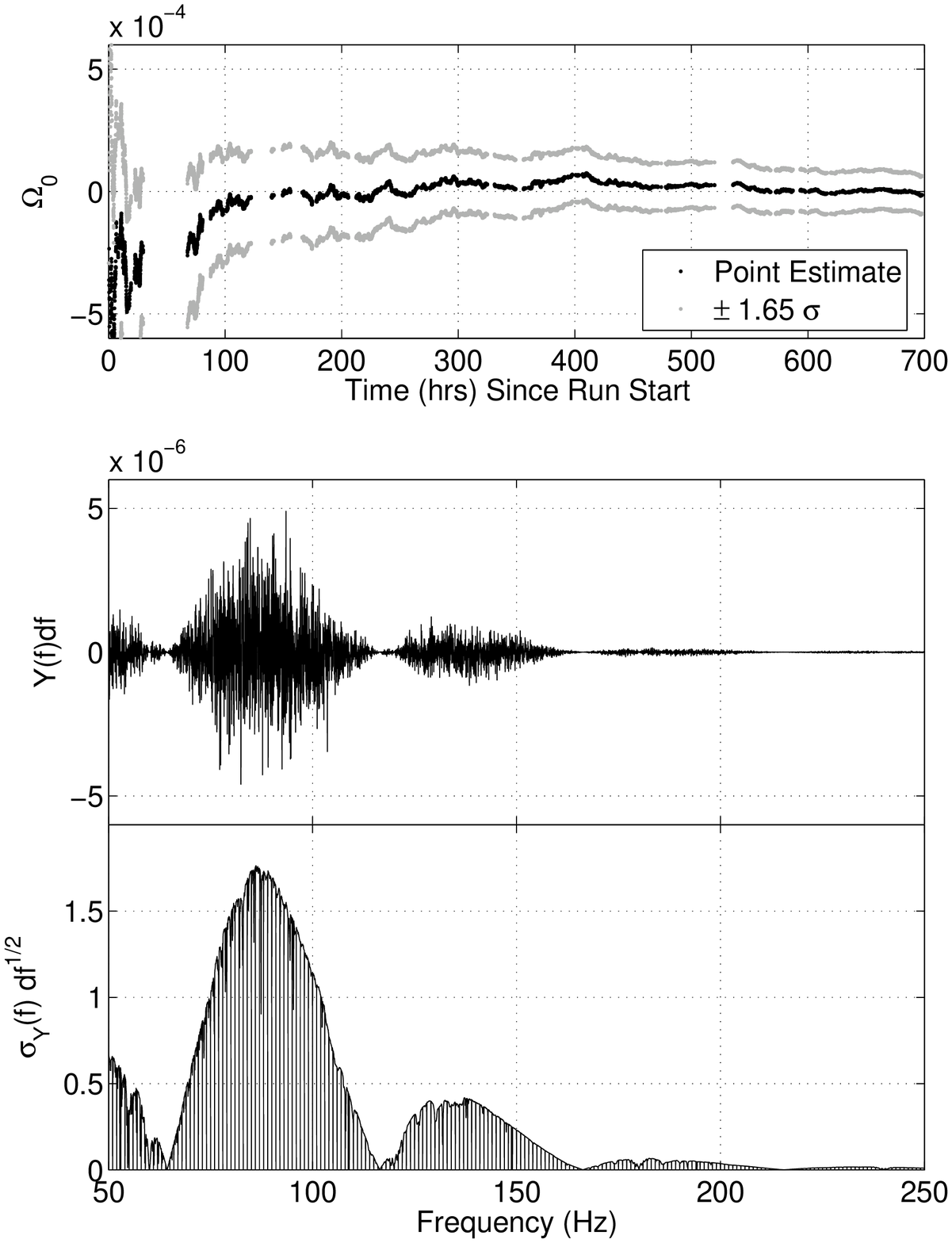}
\caption{H1L1, 192-sec analysis with $\zeta=0.3$. Top: Cummulative 
estimate of $\Omega_0$ is shown as a function
of time. Middle: cross-correlation spectrum $Y(f)$. Bottom: theoretical
uncertainty $\sigma_Y(f)$ as a function of frequency.}
\label{H1L1res}
\end{figure}

\begin{figure}
\epsscale{.80}
\plotone{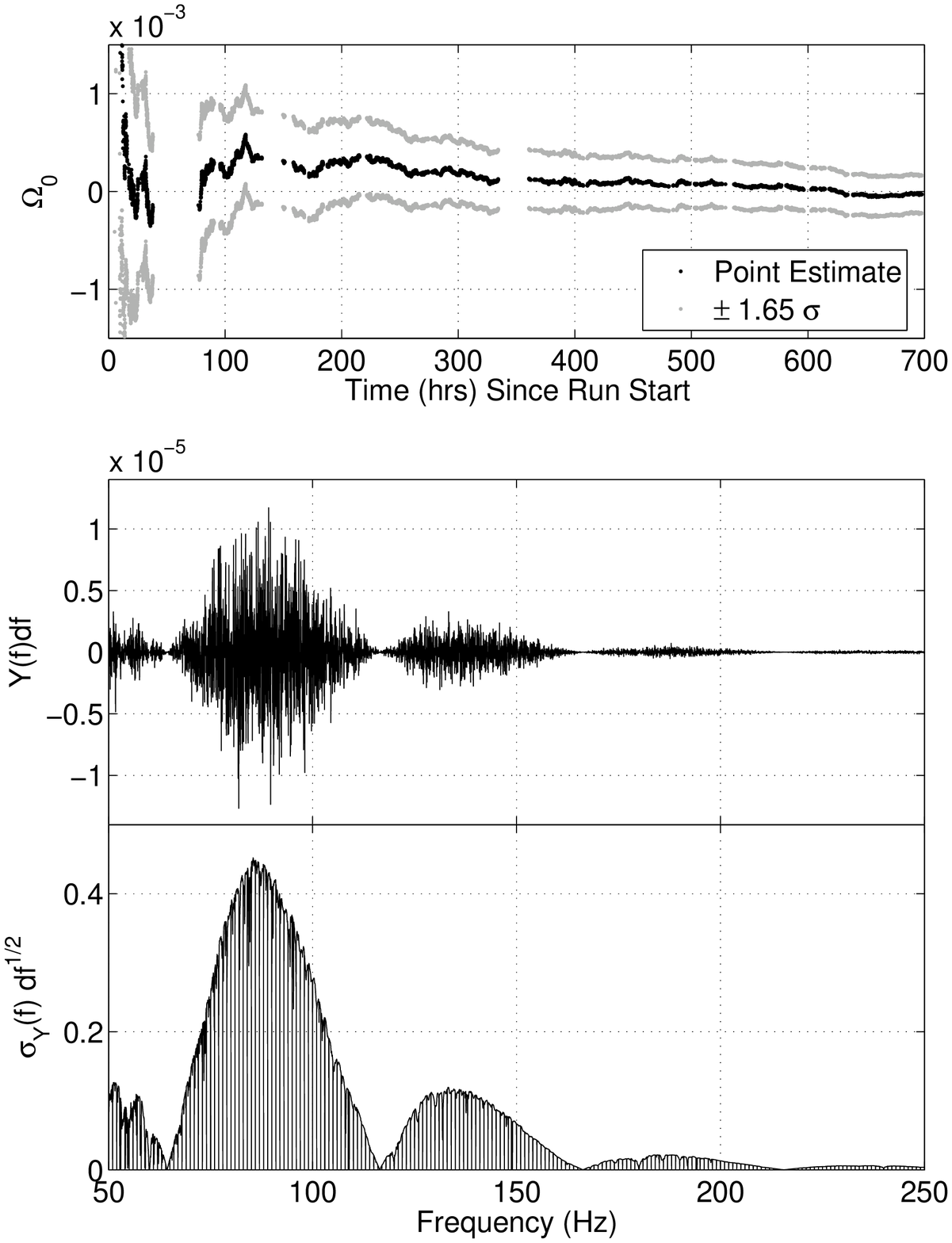}
\caption{H2L1, 192-sec analysis with $\zeta=0.3$. Top: Cummulative 
estimate of $\Omega_0$ is shown as a function
of time. Middle: cross-correlation spectrum $Y(f)$. Bottom: theoretical
uncertainty $\sigma_Y(f)$ as a function of frequency.}
\label{H2L1res}
\end{figure}

Figures \ref{H1L1res} and \ref{H2L1res} show the results of the 192-sec 
analysis with $\zeta=0.3$ for H1L1 and H2L1 respectively. The top plots in
these two Figures show the cummulative estimates for the 
frequency-independent GW template ($\alpha=0$) as a function of time. 
They indicate that there is no particular time during the run that 
dominates the result. Moreover, the $\pm 1.65\sigma$ bounds converge as
$\sim 1/\sqrt{T}$, as expected. The middle plots of the two Figures show the 
cross-correlation spectra (i.e. the integrand of Equation \ref{ptest}). 
They indicate that there is no particular frequency that dominates the 
result. 

\begin{figure}
\epsscale{.80}
\plotone{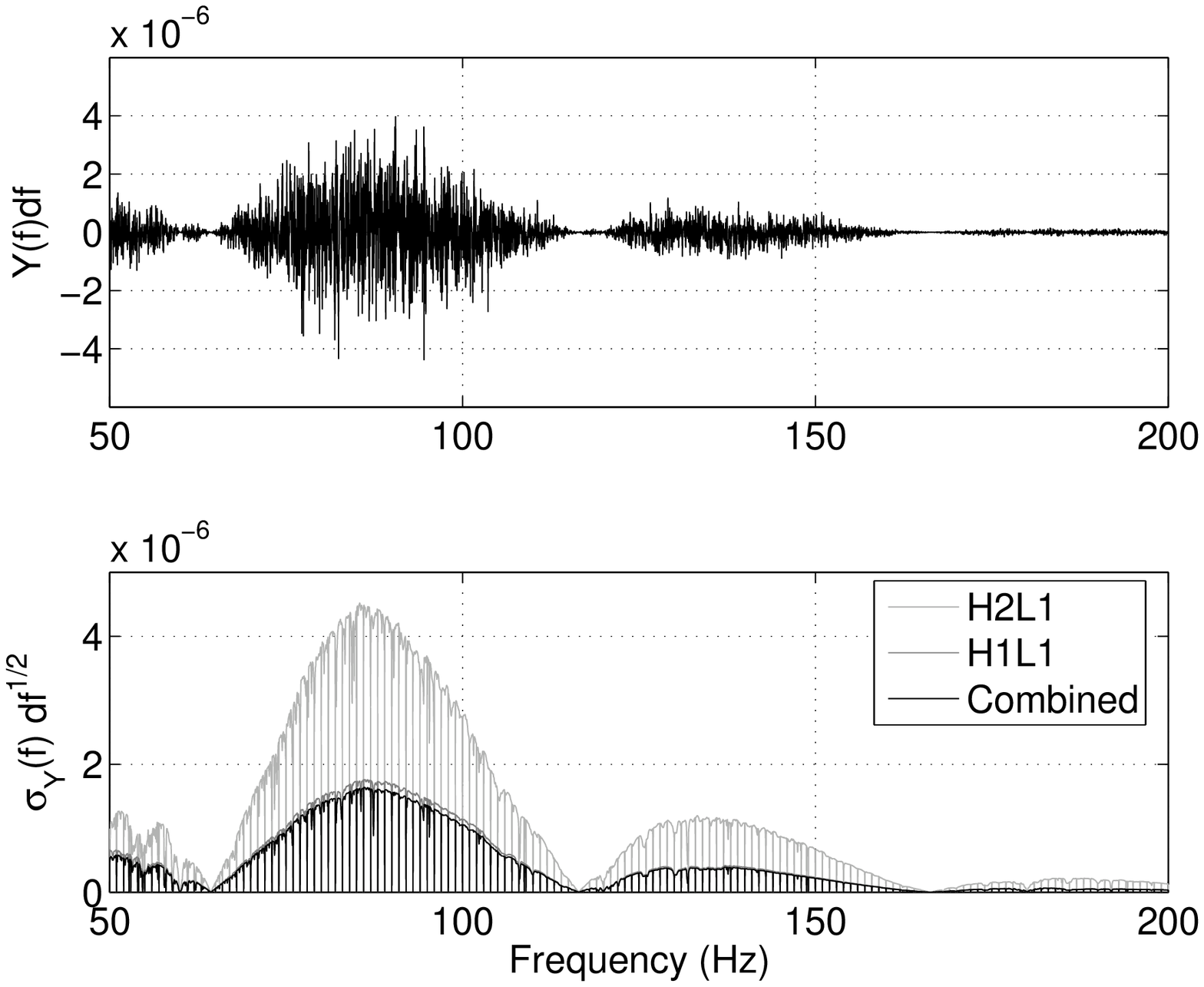}
\caption{Combined H1L1 + H2L1 result, 192-sec analysis with $\zeta=0.3$. 
Top: Combined cross-correlation spectrum. 
Bottom: Theoretical uncertainty $\sigma_Y(f)$.}
\label{finresult}
\end{figure}

We perform a weighted average of the H1L1 and H2L1 results at each frequency 
bin (with inverse variances as weights).
The resulting cross-correlation spectrum and the theoretical error are
shown in Figure \ref{finresult}. The frequency range 51-150 Hz 
contributes 99\% of the full sensitivity, as determined by the inverse 
combined variance. Integrating over this frequency range, we obtain
the final estimate for the frequency-independent spectrum:
$\Omega_0 = (-0.8 \pm 4.3) \times 10^{-5}$. 

The dominant systematic uncertainty of this result comes from the 
amplitude calibration uncertainty of the interferometers. This uncertainty
is estimated to be 5\% (L1) and 8\% (H1 and H2), and it is 
frequency-independent. The uncertainty in the
phase of the interferometer strain response is negligible compared 
to the magnitude and statistical uncertainties. Similarly,
the effect of timing errors, measured to be $\lesssim 4 {\rm \; \mu s}$, is
negligible. Using
hardware injections, we estimate that the effect of the timing errors
on our point estimate is $\lesssim 0.2$\%.

We then construct the 
Bayesian posterior distribution for $\Omega_0$ using the above
estimate, following \citep{loredo}. 
We assume a Gaussian distribution for the amplitude calibration
uncertainty (with mean 1 and standard deviation $\sqrt{0.05^2+0.08^2}=0.093$),
and we marginalize over it. We assume the prior distribution for 
$\Omega_0$ to be the posterior distribution obtained in our previous
analysis of the S3 data \citep{S3paper}. The 90\%
upper limit is the value of $\Omega_0$ for which 90\% of the 
posterior distribution lies between 0 and the upper limit.
This procedure yields the 
Bayesian 90\% UL on $\Omega_0$ of $6.5 \times 10^{-5}$. This is an improvement
by a factor $13\times$ over the previous result in the same frequency band,
established based on the science run S3 \citep{S3paper}. To 
investigate robustness of our result under different priors, we repeated
the calculation using flat priors for the amplitude calibration uncertainties,
and using a flat prior for $\Omega_0$ between 0 and $8.4\times10^{-4}$
(previous 90\% UL by LIGO, \citep{S3paper}). We found that these different
choices of priors have less than 3\% effect on the 90\% upper limit.

\begin{figure}
\epsscale{.60}
\plotone{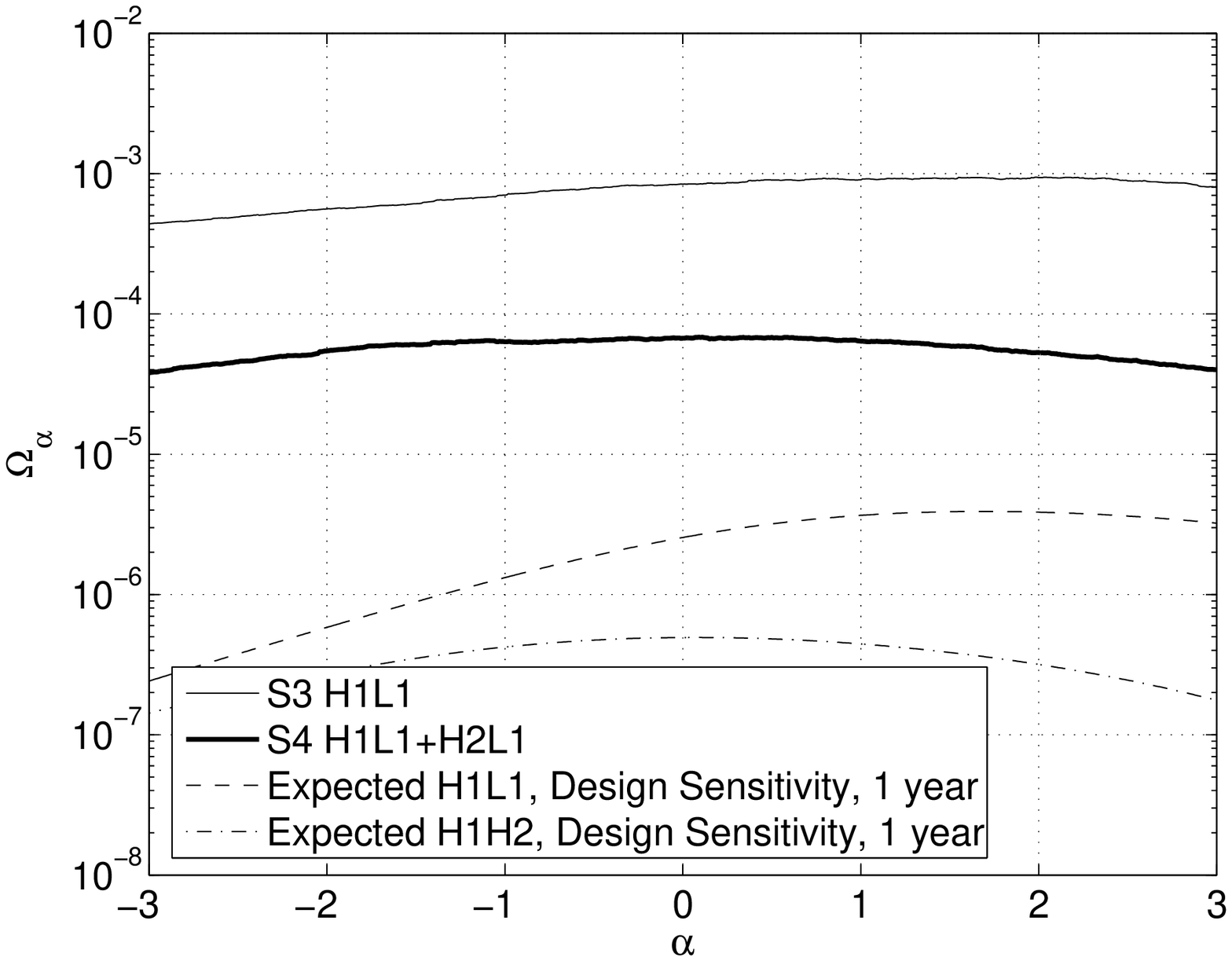}
\plotone{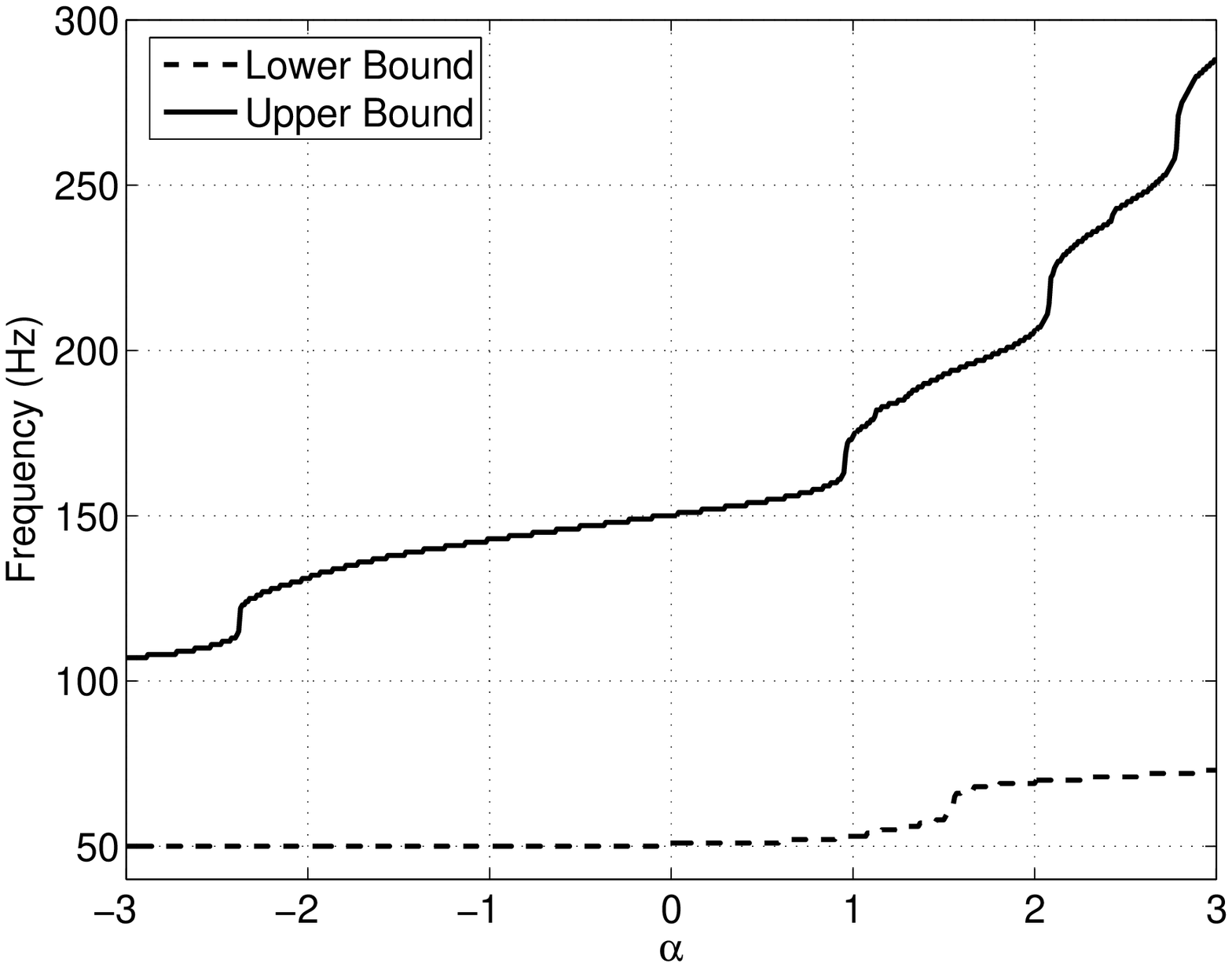}
\caption{Top: 90\% UL on $\Omega_{\alpha}$ as a function of 
$\alpha$ for S3 H1L1
and S4 H1L1+H2L1 combined, and expected final sensitivities 
of LIGO H1L1 and H1H2 pairs, assuming LIGO design sensitivity and one year
of exposure. Bottom: Frequency band containing 99\% of the full sensitivity
(as determined by the inverse variance) is plotted as a function of 
$\alpha$ for the S4 result.}
\label{ulvsalpha}
\end{figure}

Once the estimate is made for the frequency-independent spectrum, one 
can perform appropriate
frequency-dependent scalings of $Y(f)$ and $\sigma_Y(f)$, recalculate
the posterior distributions, and re-marginalize to obtain upper limits for
other templates, such as
the power-law templates with different spectral indices $\alpha$
(see Equation \ref{template}).
Figure \ref{ulvsalpha} shows the 90\% UL as a function of the
spectral index $\alpha$ obtained for this analysis. 
Similar results for
the S3 run of LIGO, as well as the expected sensitivities of H1L1 and H1H2
pairs assuming 1 year of exposure and design interferometer sensitivities,
are also shown. 
The frequency range of interest is defined to include 99\%
of the full sensitivity, as determined by the inverse variance. For 
the S4 result, the frequency range varies 
between $(50-107)$ Hz for $\alpha = -3$ and $(73-288)$ Hz for $\alpha = +3$, 
as shown in the bottom plot of Figure \ref{ulvsalpha}. 
Note that the expected sensitivity of the collocated 
Hanford interferometer pair (H1H2) is
significantly better than that of the H1L1 pair due to the better 
antenna pattern overlap (see Figure \ref{overlap}). However, this pair 
is also more susceptible to 
instumentally correlated noise. New analysis methods are being pursued 
to estimate and suppress these instrumental correlations \citep{nickH1H2}.

\subsection{Signal Injections}
\begin{figure}
\epsscale{.80}
\plotone{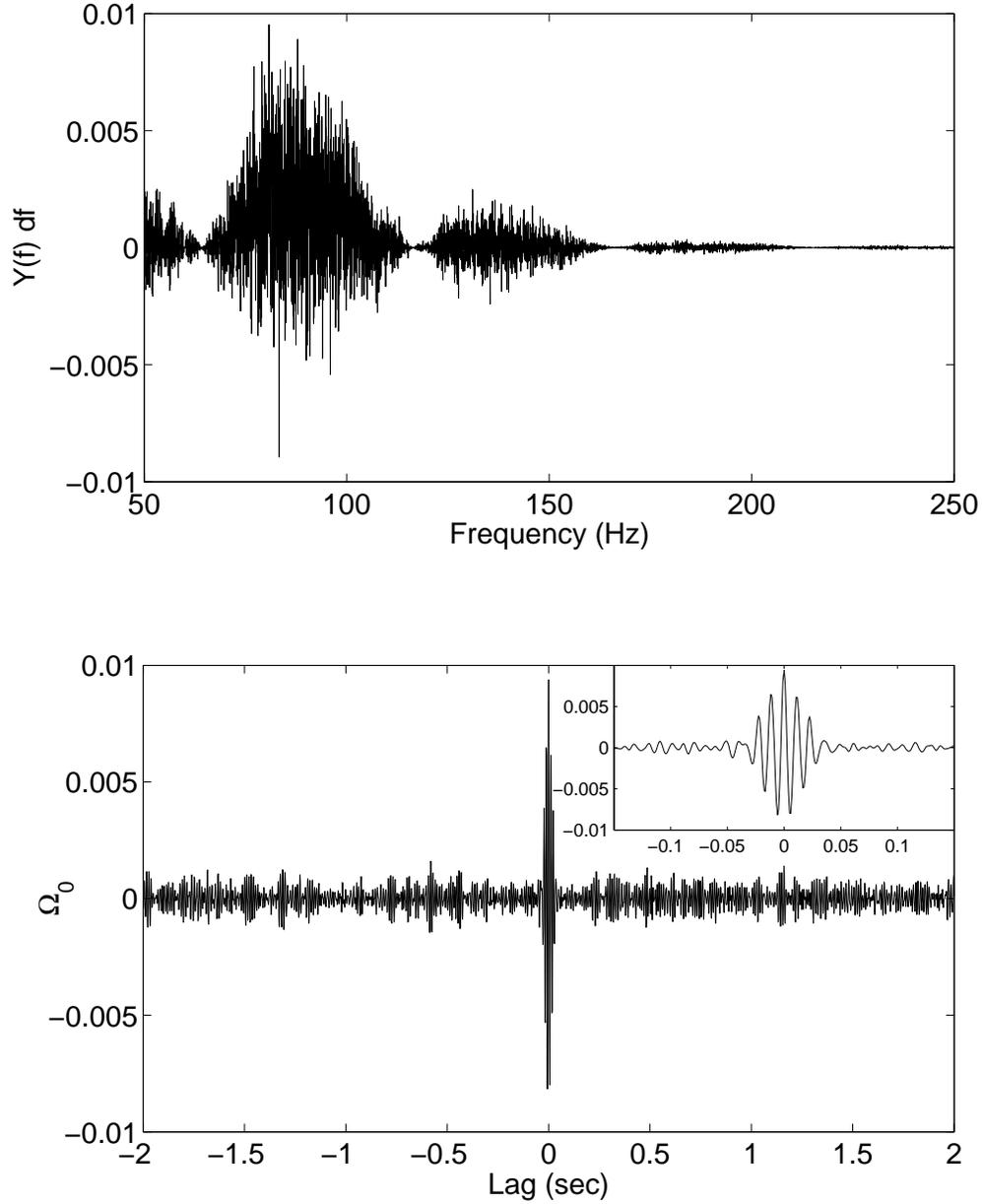}
\caption{H1L1 injection 3 
(with intended injection amplitude $\Omega_0 = 1.1 \times 10^{-2}$). 
Top: Cross-correlation spectrum $Y(f)$. Bottom: Inverse
Fourier transform of the cross-correlation spectrum indicates the clear 
signal at zero lag. The inset plot is a zoom-in around zero lag.}
\label{inj4}
\end{figure}

We exercise the analysis procedure described above using simulated
stochastic signals injected into the data both in software and in
hardware - see \citep{Bose:2003nb}. In particular, we verify that
the recovery of the injected signals is not affected by the 
data-quality cuts we impose.
The hardware injections are performed by physically moving the mirrors to
simulate a stochastic GW signal. Three hardware injections
were performed during the S4 run, 
all using a frequency-independent GW spectrum ($\alpha=0$).
Table \ref{hwinj} summarizes the recovery of all of the hardware injections.
Figure \ref{inj4} shows the cross-correlation spectrum for injection 3.
It also shows the inverse Fourier transform of the spectrum, which 
is equivalent to the estimate of $\Omega_0$ for different values of 
time-lag between two interferometers (for short time-lags).

\begin{table*}[!th]
\centering
\begin{tabular}{|r|r|r|r|r|}
\hline
Injection & H1L1 Expected & H1L1 Recovered & H2L1 Expected & H2L1 Recovered\\
 & $(\times 10^{-2})$ & $(\times 10^{-2})$ & $(\times 10^{-2})$ & 
$(\times 10^{-2})$ \\
\hline
1 & 9.1 & $7.9 \pm 0.2 \pm 0.8$ & 7.9 & $6.9 \pm 0.4 \pm 0.6$ \\
2 & 2.5 & $2.5 \pm 0.4 \pm 0.2$ & 2.3 & $1.5 \pm 0.6 \pm 0.2$\\
3 & 1.1 & $0.95 \pm 0.04 \pm 0.10$ & - & - \\
\hline
\end{tabular}
\caption{Summary of the hardware injection amplitudes during S4. 
The second and fourth columns
indicate the expected injection amplitudes for H1L1 and H2L1 respectively, 
based on the signal injected 
into the differential-arm servo. The third and the fifth
columns list the recovered values using the H1L1 and H2L1 pairs respectively. 
The recovered values are listed with statistical errors (as defined
in Equation \ref{sigma}), and with systematic errors (estimated
using 5\% calibration uncertainty in L1 and 8\% calibration 
uncertainty in H1 and H2, added in quadrature).
For injection 3, the data of H2 interferometer were compromised due to a
failure of the interferometer's laser.}
\label{hwinj}
\end{table*}

\begin{figure}
\epsscale{.80}
\plotone{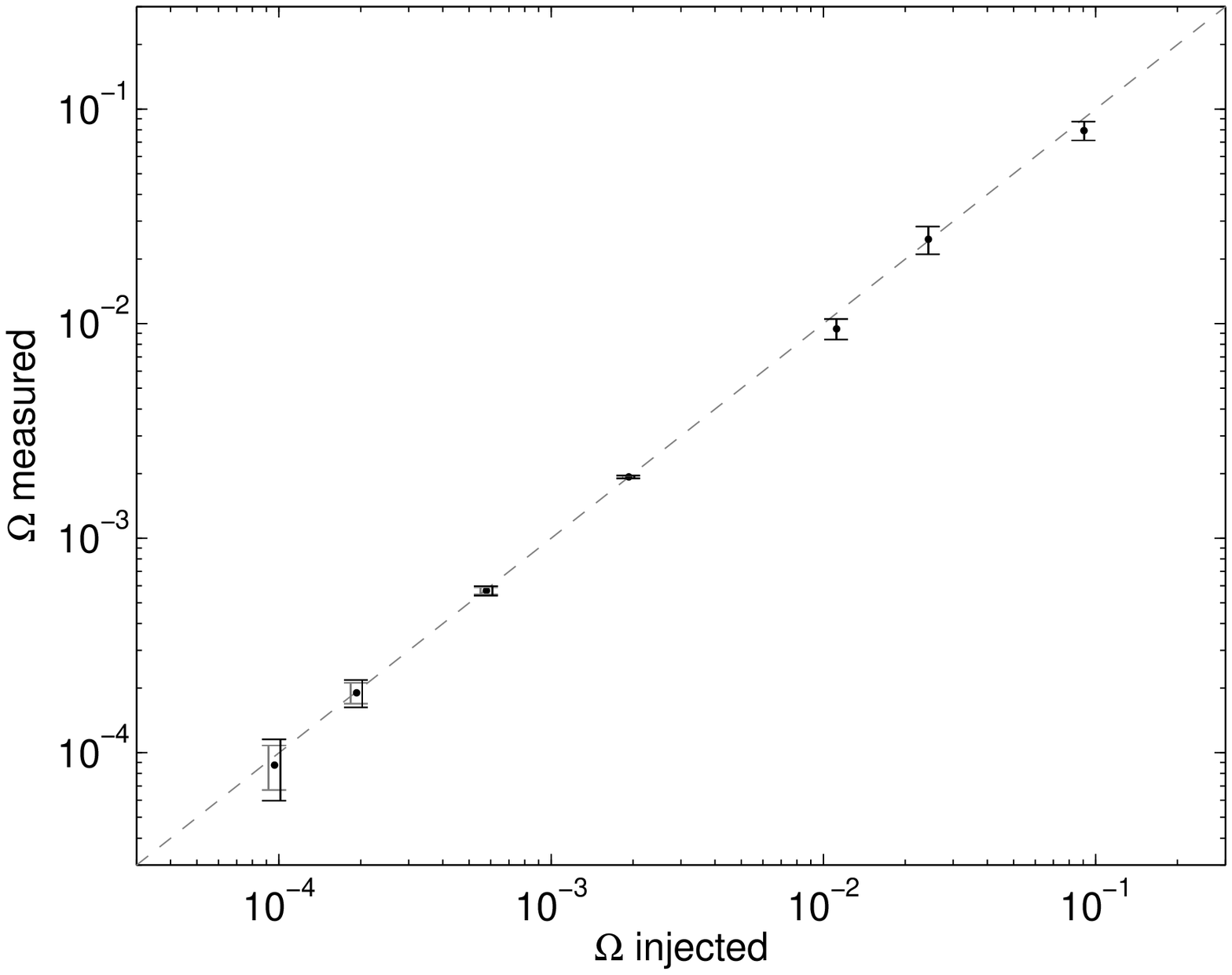}
\caption{Injections using H1L1 data: 10 trials were performed
for software injections with amplitudes 
$\Omega_0 = 1\times 10^{-4}$, $2\times 10^{-4}$,
$6\times 10^{-4}$, and $2\times 10^{-3}$. The left(gray) error bars
denote the theoretical errors, while the right(black) error bars denote the
standard errors over the 10 trials. The remaining points correspond to
the three hardware injections listed in Table \ref{hwinj}; their
error bars correspond to statistical and systematic errors added in 
quadrature, as shown in Table \ref{hwinj}.}
\label{softinj}
\end{figure}

We performed a sequence of software injections, where the injected
signal is simply added to the interferometer data in the analysis. We
performed 10 trials for 4 injection amplitudes using about $1/3$ of the
S4 H1L1 data. Figure \ref{softinj} shows that the signal is successfully
recovered down to $1 \times 10^{-4}$ in $\Omega_0$. Moreover, the 
theoretical error bars agree well with the standard deviation over the 10
trials. All injections were performed assuming a frequency-independent
GW spectrum ($\alpha=0$).

\section{Implications}

In this Section, we investigate the implications of the new upper limit
for some of the models of the stochastic GW background. We also discuss
the complementarity of our result with various other
experimental constraints on the stochastic GW background. 

\subsection{Complementarity with Other Measurements and Observations}
\begin{figure}
\epsscale{0.8}
\plotone{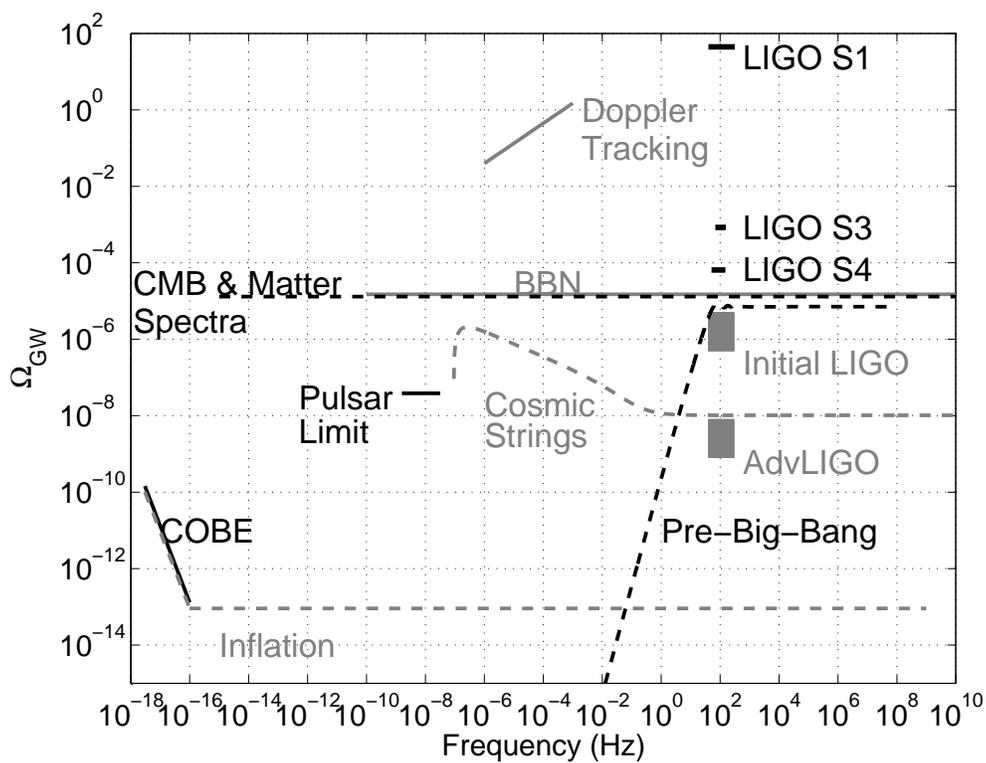}
\caption{Landscape plot (see text for details). The curves corresponding
to inflationary, cosmic-string, and pre-big-bang models are examples;
significant variations of the predicted spectra are possible as the 
model parameters are varied. The bounds labeled ``BBN'' and ``CMB and Matter
Spectra'' apply to the integral of the GW spectrum over the frequency range
spanned by the corresponding lines.}
\label{landscape}
\end{figure}

Figure \ref{landscape} compares different experiments and some of the
theoretical models. For wavelengths larger than the horizon size
at the surface of last scattering (redshifted to today, this corresponds
to frequencies below $\sim10^{-16}$ Hz), the COBE observations of the CMB
place an upper limit on the stochastic gravitational
wave background of $\Omega_{\rm GW}(f) < 1.3 \times 10^{-13}$
\citep{cobe1}. 
In standard inflationary models~\citep{cobe2}, the GW spectrum is 
expected to be (almost) flat at frequencies above $\sim 10^{-16}$ Hz. 

The fluctuations in the arrival times of millisecond pulsar 
signals can be used to place 
a bound at $\sim 10^{-8}$ Hz~\citep{pulsar}: 
$\Omega_{\rm GW}(f) < 3.9 \times 10^{-8}$ (assuming frequency independent
GW spectrum). Similarly,
Doppler tracking of the Cassini spacecraft can be used to arrive
at yet another bound, in the $10^{-6}-10^{-3}$ Hz band
\citep{doppler}: $\Omega_{\rm GW}(f) < 0.027$. 

If the energy density carried by the gravitational waves at the time of
Big-Bang Nucleosynthesis (BBN) were large, the amounts of the
light nuclei produced in the process could be altered. Hence, the BBN
model and observations can be used to constrain the total energy carried 
by gravitational waves at the time of 
nucleosynthesis~\citep{BBN,maggiore,allen}: 
\begin{equation}
\int \Omega_{\rm GW}(f) \; d(\ln f) < 1.1 \times 10^{-5} \; (N_{\nu} - 3),
\end{equation}
where $N_{\nu}$ is the effective 
number of relativistic species at the time of BBN.
Measurements of the light-element abundances, combined with the WMAP data,
give the following 95\% upper bound: $N_{\nu} - 3 < 1.4$ ~\citep{cyburt}.
This limit translates into 
$\int \Omega_{\rm GW}(f) \; d(\ln f) < 1.5 \times 10^{-5}$. This bound
applies down to $\sim 10^{-10}$ Hz, corresponding to the horizon size at
the time of BBN. 

Gravitational waves are also expected to leave a possible imprint on the 
CMB and matter spectra, similar
to that of massless neutrinos. \citep{smith} used recent measurements
of the CMB anisotropy spectrum, galaxy power spectrum, and of the
Lyman$-\alpha$ forest, 
to constrain the energy density carried by gravitational waves to 
$\int \Omega_{\rm
GW}(f) \; d(\ln f) < 1.3 \times 10^{-5}$ for homogeneous initial
conditions. This bound is competitive with the BBN bound and it extends 
down to $\sim 10^{-15}$ Hz, corresponding to the horizon size at the time
of CMB decoupling. It is also expected to improve as
new experiments come online (such as Planck or CMBPol).


The LIGO results apply to the frequency region around 100 Hz. The result 
discussed in this paper is an improvement by a factor 
$13\times$ over the previous LIGO result in the 100 Hz region, 
for a frequency-independent spectrum of GW background. 
A one-year run at design sensitivity of LIGO (the S5 run, 
which began in November of 2005)
is expected to improve the sensitivity by another factor
$10\times$ - $100\times$, while
Advanced LIGO is expected to achieve sensitivities better by yet another
factor $100\times$ - $1000\times$, 
eventually reaching $10^{-9}-10^{-8}$ for $\Omega_0$. The uncertainty in the
final reach of LIGO and of Advanced LIGO comes from the potential 
instrumental correlations that could be present between the
collocated Hanford interferometers and from the uncertainty in the
final optical configurations of interferometers in Advanced LIGO.
The result discussed in this paper is still weaker than the indirect BBN 
bound, but the future runs by LIGO and Advanced LIGO are expected
to surpass this bound.

The standard inflationary models are most stringently constrained by the
CMB bound at lowest frequencies. Although they are most likely out of 
range of LIGO and of Advanced LIGO, they may be accessible to future
GW interferometers \citep{smith2}. 
However, there are models of stochastic GW background
that LIGO is beginning to explore. We illustrate this with examples of
cosmic string and pre-big-bang models. 

\subsection{Implications for Cosmic Strings Models}

Cosmic strings can be formed as linear defects during symmetry breaking
phase transitions in the early Universe as well as in string theory
inspired inflation scenarios. In the latter case they have been dubbed
cosmic superstrings. CMB data is not consistent with cosmic strings as the
predominant source of density fluctuations in the Universe. Their
existence, however, is not ruled out below the GUT scale and cosmic
strings may still lead to a myriad of detectable astrophysical signatures
such as gravitational radiation, gamma-ray bursts and ultra-high energy
cosmic rays. For a review, see \citep{vilenkinbook}.

\citep{CS3} investigated the stochastic background of
gravitational waves produced by cusps on cosmic strings, integrated
over all redshifts and all directions on the sky. They find that three 
parameters define the gravitational-wave spectrum
due to cosmic strings:
\begin{itemize}
\item String tension $\mu$: This parameter is usually expressed as a 
dimensionless quantity $G\mu$ (assuming speed of light $c=1$), 
where $G$ is Newton's constant. String-theory
inspired inflation scenarios 
prefer the range $10^{-11} \le G\mu \le 10^{-6}$. 
\item Reconection probability $p$: While ordinary, field-theoretic 
strings reconnect whenever they intersect ($p=1$), 
the reconnection probability for
superstrings is typically smaller than 1. In particular, the theoretically
favored range is $10^{-3} \le p \le 1$ \citep{jackson}. 
\item $\epsilon$: This parameter describes the typical
size of the closed loops produced in the string network. The value of this
parameter is uncertain, and it can span several orders of magnitude.
We will consider the range $10^{-13} < \epsilon < 10^{-2}$, which is both
theoretically viable and most interesting from the point of view of LIGO. 
\end{itemize}

These parameters determine both the amplitude and the shape of the 
gravitational-wave spectrum. In particular, parameters $\epsilon$ and 
$G\mu$ determine the lowest frequency (at a given redshift) 
at which a string loop could emit gravitational radiation. 
Since there is a low-frequency cut-off to the predicted 
gravitational-wave spectrum, it is
possible to have a cosmic string model that would avoid the low-frequency
bounds due to CMB or pulsar timing measurements, but still be within
reach of LIGO (c.f. Figure \ref{landscape}). 

\begin{figure}
\epsscale{0.8}
\plotone{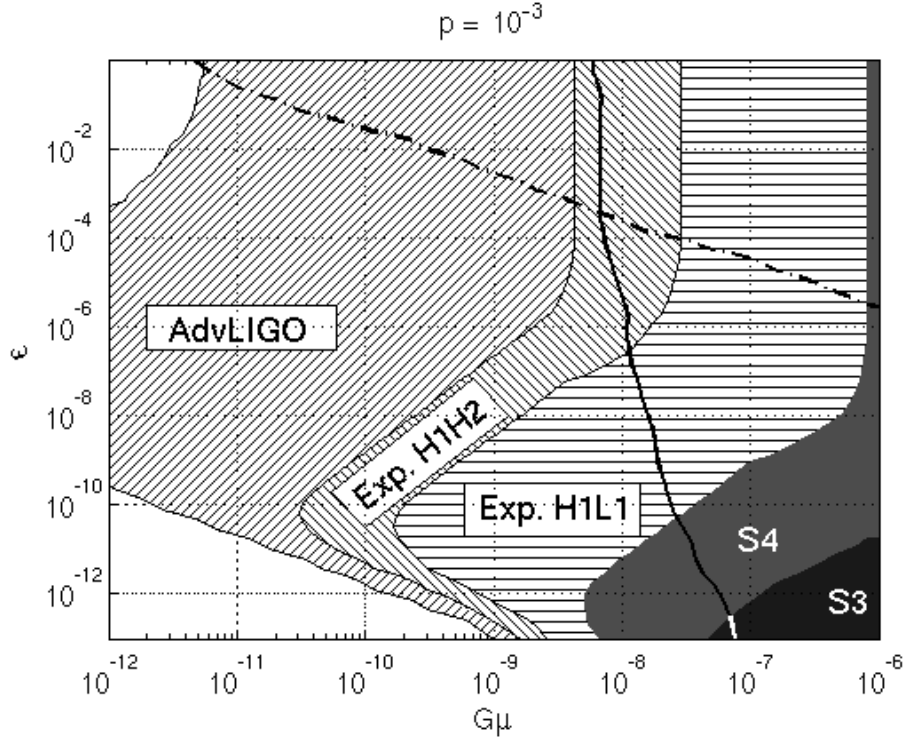}
\caption{The $\epsilon$ - $G\mu$ plane for the cosmic string models with 
$p = 10^{-3}$. The shaded regions are excluded by the LIGO S3 limit (darker)
and by the LIGO S4 limit presented here (lighter). The hatched regions
are accessible to future LIGO runs:
expected LIGO sensitivity for the H1L1 pair, 
assuming design interferometer strain sensitivity, and 1 year of exposure
('$-$'); 
expected LIGO sensitivity for the H1H2 pair, assuming 
design interferometer strain sensitivity, and 1 year of exposure 
('$\backslash$'); 
expected Advanced LIGO sensitivity for the H1H2 pair, 
assuming interferometer strain sensitivity tuned for the binary neutron star 
inspiral search and 1 year of exposure ('/').
The dash-dotted black curve is the exclusion curve based on the pulsar limit 
\citep{pulsar} (the excluded region is above the curve). The solid black
curve is the exclusion curve based on the indirect big-bang-nucleosynthesis 
bound (the excluded region is to the right of the curve).
}
\label{epsvsGmu}
\end{figure}

Figure \ref{epsvsGmu} shows the region of the parameter space (for
$p=10^{-3}$) excluded by the result discussed in this paper and by the S3 
result \citep{S3paper}, as well as 
the expected reach of LIGO and Advanced LIGO in future runs. 
As shown in this Figure, LIGO is most sensitive to the regions of 
large $G\mu$ and small $\epsilon$. Moreover, LIGO results are
complementary to the pulsar timing limit, which is 
most sensitive to models with large $G\mu$ and large $\epsilon$, 
and to the indirect BBN limit. In particular, the population of models
with $p=10^{-3}$, $\epsilon \lesssim 5 \times 10^{-11}$, and
$5 \times 10^{-9} \lesssim G\mu \lesssim 7 \times 10^{-8}$ 
are excluded by the result discussed in this paper, but are not
accessible to other current experimental bounds.
To produce this Figure, we used Equations 4.1-4.7 from \citep{CS3}.

We note that as $p$ increases toward 1, the bounds/reach of all 
experiments in the $G\mu-\epsilon$ plane weaken because
the amplitude of the gravitational-wave spectrum scales with $1/p$. We also
note that there are significant uncertainties in the calculation discussed by 
\citep{CS3}, mostly due to incomplete understanding of the string 
network behavior. 
For example, the size of loops is usually assumed to be given by the
scale of gravitational back reaction. Recent numerical simulations of
cosmic string networks \citep{MartinsShellard}, \citep{Vanchurin}, and 
\citep{Ringeval} instead suggest that loop sizes are related to the
large scale dynamics of the network. If this is the case then the expected
stochastic background at the frequencies of pulsar timing experiments
could be subtantially larger \citep{Hogan}. Additionally, the treatment of
\citep{CS3} did not account for the effects of late-time
acceleration. Thus a more detailed analysis of the stochastic background
produced by cosmic strings is necessary.

\subsection{Implications for Pre-Big-Bang Models}

Pre-big-bang models are cosmology models motivated by string theory 
\citep{pbb}, \cite{pbbrep}. In these models, the Universe evolves through
several phases: ``dilaton'' phase in which the Universe is large and 
shrinking; ``stringy'' phase in which the curvature of the Universe is 
high; and standard radiation and matter dominated phases.
The GW spectrum is generated by amplification of vacuum fluctuations
as the Universe transitions from one phase to another. The shape and 
amplitude of the
spectrum are determined by the states of the Universe in the different phases. 
Although the ``stringy phase'' of the model and the transition
to the radiation phase are not well understood, some models have been
proposed in the literature that may partially describe it. In the formalism
developed by \citep{BMU}, the GW spectrum produced by the model can be
described as:
\begin{itemize}
\item $\Omega_{\rm GW}(f) \sim f^3$ for $f<f_s$, where $f_s$ is essentially
unconstrained.
\item $\Omega_{\rm GW}(f) \sim f^{3-2\mu}$ for $f_s < f < f_1$, where
$\mu<1.5$ defines the evolution of the Universe in the ``stringy'' phase.
The cutoff frequency $f_1$ is determined by string-related parameters 
and its most natural
value is expected to be $4.3 \times 10^{10}$ Hz. 
\end{itemize}

\begin{figure}
\epsscale{0.8}
\plotone{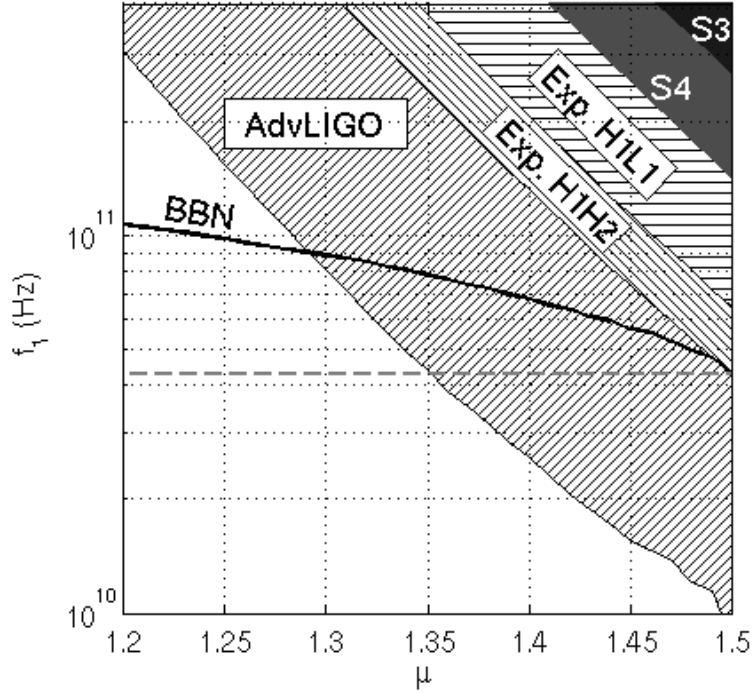}
\caption{The $f_1-\mu$ plane for the pre-big-bang models with $f_s=30$ Hz. 
The shaded regions are excluded by the LIGO S3 limit (darker)
and by the LIGO S4 limit presented here (lighter). The hatched regions
are accessible to future LIGO runs:
expected LIGO sensitivity for the H1L1 pair, 
assuming design interferometer strain sensitivity, and 1 year of exposure
('$-$'); 
expected LIGO sensitivity for the H1H2 pair, assuming 
design interferometer strain sensitivity, and 1 year of exposure 
('$\backslash$'); 
expected Advanced LIGO sensitivity for the H1H2 pair, 
assuming interferometer strain sensitivity tuned for the binary neutron star 
inspiral search and 1 year of exposure ('/').
The solid black curve is the exclusion curve 
based on the BBN limit (the excluded region is above the curve). The
horizontal dashed line denotes the most natural value of 
$f_1 = 4.3 \times 10^{10}$ Hz.}
\label{f1vsmu}
\end{figure}

Following the analysis in \citep{MB}, we scan the parameter space 
$(f_s,\mu,f_1)$. For each set of values of these parameters, we 
calculate the GW spectrum following \citep{BMU} and check whether the
model is accessible to the current (and past) LIGO results. We also
project the sensitivity of initial and Advanced LIGO to these models.
Figure \ref{f1vsmu} shows the $f_1-\mu$ plane for $f_s = 30$ Hz. Note that
the LIGO S3 and S4 results are beginning to explore
the parameter space of these models, although the indirect BBN bound is
still a stronger constraint. Future runs of LIGO and 
of Advanced LIGO are expected to explore significantly larger
parts of the parameter space, eventually surpassing the BBN bound (in
some parts of the parameter space) and even reaching the most
natural value of $f_1 = 4.3 \times 10^{10}$ Hz.

\section{Conclusions}

LIGO data acquired during the
science run S4 yield a new Bayesian 90\% upper limit on the 
amplitude of the stochastic GW background: $\Omega_0 < 6.5 \times 10^{-5}$
for the frequency-independent GW spectrum ($\alpha=0$) in the 
frequency band 51-150 Hz. 
Similar limits are obtained for
other values of $\alpha$, as shown in Figure \ref{ulvsalpha}. This result
is an improvement by a factor 13$\times$ over the previous upper limit
in the same frequency range, obtained by LIGO in the science run S3. 

This result is obtained using 192-sec long intervals of data
with 1/32 Hz frequency resolution of the spectra, and it properly
excludes the known instrumental correlations at 1 Hz harmonics. It is 
fully consistent with the blind result that uses 60-sec long 
intervals of data with 1/4 Hz frequency resolution, which is slightly 
contaminated by the instrumental 1 Hz harmonics. 
It is also more conservative than the blind result, as the theoretical error
is larger due to the smaller amount
of data available in the form of acceptable 192-sec intervals 
(as compared to the 60-sec intervals).

This result is complementary to the constraints on the gravitational-wave 
spectrum, based on the measurements of the CMB spectrum 
and pulsar timing. It is still weaker than the indirect BBN bound in 
the relevant 
frequency range. The ongoing 1-year long run of LIGO at the design
sensitivity, and the future runs of Advanced LIGO, are expected to
surpass the BBN bound. Furthermore, this result is already exploring
the parameter space of some models of the stochastic GW background, such
as cosmic strings models and pre-big-bang models. 

The authors gratefully acknowledge the support of the United States
National Science Foundation for the construction and operation of
the LIGO Laboratory and the Particle Physics and Astronomy Research
Council of the United Kingdom, the Max-Planck-Society and the State
of Niedersachsen/Germany for support of the construction and
operation of the GEO600 detector. The authors also gratefully
acknowledge the support of the research by these agencies and by the
Australian Research Council, the Natural Sciences and Engineering
Research Council of Canada, the Council of Scientific and Industrial
Research of India, the Department of Science and Technology of
India, the Spanish Ministerio de Educacion y Ciencia, The National
Aeronautics and Space Administration, the John Simon Guggenheim
Foundation, the Alexander von Humboldt Foundation, the Leverhulme
Trust, the David and Lucile Packard Foundation, the Research
Corporation, and the Alfred P. Sloan Foundation. LIGO DCC number: P060012-06-D.

{}

\end{document}